%% file: main.tex
\documentclass{article}
\usepackage[utf8]{inputenc}
\usepackage{standalone}
\usepackage{subfigure}
\usepackage{adjustbox}
\usepackage[fontsize=13pt]{scrextend}
\usepackage[english]{babel}   
\usepackage{tikz}
\usetikzlibrary{shapes}
\usetikzlibrary{arrows}
\usepackage{amsmath}  
\usepackage{amsthm}
\usepackage{amsfonts}
\usepackage{biblatex}
\usepackage[hidelinks]{hyperref}
\usepackage{geometry}
\usepackage{csquotes}
\title{Stack number and queue number of graphs}
\author{Adam Straka}
\date{February 2023}

\begin{document}

\theoremstyle{definition}
\newtheorem{case}{Case}
\newtheorem{theorem}{Theorem}[subsection]
\newtheorem{lemma}[theorem]{Lemma}

\newtheorem{definition}[theorem]{Definition}

\newenvironment{sketch_proof}{
\renewcommand{\proofname}{Proof sketch}\proof}{\endproof}

\maketitle
\begin{abstract}
     In this paper we give an overview of the graph invariants queue number and stack number (the latter also called the page number or book thickness). Due to their similarity, it has been studied for a long time, whether one of them is bounded in terms of the other. It is now known that the stack number is not bounded by the queue number. We present a simplified proof of this result. We also survey the known results about possible stack number bound on the queue number. This preprint is a rework of the bachelor thesis~\cite{Straka2023thesis}.
\end{abstract}
\newpage
\section{Introduction}

Our work is motivated by the task of so-called linearization. Linearization of a graph consist of "stretching" a graph into a line and processing its edges usually using several stacks or queues. The goal is to use as few stacks or queues as possible. The question is, which of the two data structures is better at linearizing graphs?

Stack number and queue number are graph invariants which could decide the relative power of their underlying data structures in laying graphs in lines. This approach was proposed by Heath, Leighton, and Rosenberg in ~\cite{comparing}. We define them formally in Chapters \ref{chapter_stack} and \ref{chapter_queue}.

Stack number was introduced by Atneosen and Persinger in the sixties ~\cite{persinger1966subsets, atneosen1968embeddability}. It was first defined as a problem of embedding a graph into a book with its vertices in the back of the book and edges on the pages of the book. Hence the name page number or book thickness. 

Queue number was introduced later, in 1992 by Heath and Rosenberg ~\cite{laying_queues}. It is relevant to several algorithms which are based on queues. However, its relation to stack number was immediately a focal point of research regarding queue number.


\subsection{Outline}
In Chapters \ref{chapter_stack} and \ref{chapter_queue}, we give an overview of the stack number and the queue number. We look at the complexity of determining the invariants and present some classes of given stack number or queue number. We also look at the queue number and the stack number of some classes of graphs.

In Chapter \ref{chapter_oposite}, we summarize what is known about the stack number bound on the queue number. Furthermore, we determine the queue number of $1$ and $2$-stack graphs.

In Chapter \ref{chapter_proof}, we prove that the stack number is not bounded by the queue number.

\subsection{Preliminaries}
We list all the necessary definitions and theorems.
\subsubsection{Order}
\begin{definition}
  (\emph{Partial order}) We say that relation $\prec$ is a partial order on a set $A$ if $\prec$ is reflexive, antisymmetric, and transitive.
\end{definition}

\begin{definition}
  (\emph{Linear order}) We call a partial order $\prec$ on a set $A$ linear (or total) if for all $a,b \in A$, either $a \prec b$ or $b \prec a$.
\end{definition}

\subsubsection{Graph theory}
\begin{definition}
  (\emph{Simple graph}) We say that $G = (V, E)$ is a simple graph, where $V$ is a set and  $E \subseteq {\binom{V}{2}}$ is a set of 2 element subsets of $V$. We call $V(G) = V$ the vertex set of the graph $G$ and $E(G) = E$ its edge set. For different vertices $u, v \in V(G)$, we call $uv = vu$ an edge between $u$ and $v$.
\end{definition}

\begin{definition}
     
(\emph{Connected, 2-connected graph}) We say that a simple graph $G$ is:
\begin{itemize}

\item connected if there is a path between any two vertices of $G$.
\item 2-connected if after removing any one of its vertices, the resulting graph is connected.
\end{itemize}

\end{definition}

\begin{definition}
    (\emph{Subgraph}) Let $G, H$ be simple graphs. We say that $H$ is a subgraph of $G$ if $V(H) \subseteq V(G)$, $E(H) \subseteq E(G)$, and for every $uv \in E(H)$, $u \in V(H)$ and $v \in V(H)$.
    \end{definition}

\begin{definition}
    (\emph{Planar graph}) A graph $G$ is planar if it can be embedded in the plane. This means it can be drawn in the plane without any edges crossing.
\end{definition}
\begin{definition}
    (\emph{Face}) Given a planar embedding of a simple graph $G$, a face is a maximal connected region in the complement of the planar embedding.
\end{definition}

\begin{definition}
    (\emph{Outerplanar graph}) A graph $G$ is outerplanar if it has a planar drawing in which all vertices of $G$ are incident to the outer face.
\end{definition}

\begin{definition}
    (\emph{Hamiltonian cycle}) A Hamiltonian cycle of a simple graph $G$ is a cycle in $G$ that visits each vertex of $G$ exactly once.
\end{definition}

\begin{definition}
    (\emph{Hamiltonian, subhammiltonian graph}) A simple graph $G$ is:
    \begin{itemize}
    \item Hamiltonian if it contains a Hamiltonian cycle. 
    \item subhamiltonian if it is a subgraph of a planar Hamiltonian graph. 
    \end{itemize}
\end{definition}
\begin{definition}
    (\emph{Intersection graph}) A graph $G$ is an intersection graph of a finite family of non-empty sets $\mathcal{F}$ if there is a bijection between the sets of $\mathcal{F}$ and vertices of $G$, such that two vertices in $G$ are adjacent if their corresponding sets inf $\mathcal{F}$ have non-empty intersection.
\end{definition}

\begin{definition}
    (\emph{Circle graph}) Circle graphs are intersection graphs of chord diagrams.
\end{definition}

\begin{definition}
(\emph{Triangulation, quasi-triangulation}) A planar graph $G$ is:
\begin{itemize}
    
\item a triangulation if  every face of $G$ is a triangle.
\item a quasi-triangulation if every face of $G$, except the outer face, is a triangle.
\end{itemize}

\end{definition}

\begin{definition}
  (\emph{Crossing, nested edges}) Let $G$ be a simple graph and $\prec$ a linear order on $V(G)$. Let $ab$ and $cd$ be two edges of $G$ where $a \prec b$ and $c \prec d$. We say that $ab$ and $cd$ are:
  \begin{itemize}
      
  \item crossing with respect to $\prec$ if $a \prec c \prec b \prec d$ or $c \prec a \prec d \prec b$ 
  \item nested with respect to $\prec$ if $a \prec c \prec d \prec b$ or $c \prec a \prec b \prec d$ 
  \end{itemize}
  
\end{definition}
\begin{definition}
    (\emph{Chromatic number}) Let $G$ be a graph. The chromatic dumber of $G$, denoted $\chi(G) = k$ is the minimal integer $k$, such that  there exists an assignment of $k$ colors to the vertices of $G$, where no two adjacent vertices have the same color.
\end{definition}
\begin{definition}
    (\emph{Vertex cover}) Vertex cover of a graph $G$ is a subset $V'$ of $V(G)$ such that every edge in $E(G)$ has at least one endpoint in $V'$. We denote it by $vc(G) = k$, where $k$ is the minimal positive integer, such that there exists a vertex cover of $G$ of size $k$. 
\end{definition}
\begin{definition}
    (\emph{Genus}) Genus of a connected orientable surface is an integer representing the maximum number of non-intersecting simple closed curves that can be drawn on the surface without separating it.
\end{definition}
\begin{definition}
    (\emph{Genus of a graph}) Genus of a graph $G$ is the minimal integer $g$, such that $G$ can be drawn in the orientable surface of genus $g$ without any of its edges crossing.
\end{definition}

\begin{definition}
    (\emph{Dual graph}) The dual graph $G$ of a planar graph $H$ is a graph which has a vertex for every face of $G$. There is an edge between two vertices of $G$ if their corresponding faces in $H$ are adjacent.
\end{definition}

\begin{definition}
    (\emph{Cartesian product}) Cartesian product of simple graphs $G$ and $H$ is a graph $X$, where $V(X) = V(G) \times V(H)$ and two vertices $(u, v)$ and $(u', v')$ are adjacent if $u = u'$ and $vv' \in E(H)$ or $v = v'$ and $uu' \in E(G)$. We denote the Cartesian product of $G$ and $H$ by $G \square H$.
\end{definition}

\begin{definition}
    (\emph{Subdivision, $k$-subdivision}) A graph obtained from a simple graph $G$ by replacing each edge $uv \in E(G)$ with an internally disjoint path from the rest of the graph with endpoints $u$ and $v$: 
    \begin{itemize}      
    \item of length at least $1$ is called a subdivision of $G$.
    \item with exactly $k$ internal vertices is called a $k$-subdivision.
    \end{itemize}
\end{definition}

\begin{definition}
    (\emph{Complete graph}) A simple graph $G$ is complete if any two vertices $a,b \in V(G)$ are adjacent.
\end{definition}

\begin{definition}
    (\emph{Bipartite graph}) A simple graph $G$ is bipartite if $V(G)$ can be partitioned into two disjoint sets, such that no two vertices in the same set are adjacent.
\end{definition}

\begin{definition}
    (\emph{Complete bipartite graph}) A bipartite graph with parts $A$ and $B$ is complete if every vertex $a \in A$ is adjacent to every vertex $b \in B$.
\end{definition}

\begin{definition}
    (\emph{H-decomposition}) For simple graphs $H$ and $G$, an $H$ decomposition of $G$ consists of bags $\{B_x \subseteq V(G)| x \in V(H)\}$, such that for every vertex $v \in G$ the set $\{x \in V(H) | v \in B_x\}$ induces a non-empty connected subgraph of $H$ and every edge of $G$ is contained within some bag.
\end{definition}

\begin{definition}
    (\emph{Tree-decomposition}) An $H$-decomposition is called a tree-decomposition if $H$ is a tree.
\end{definition}
\begin{definition}
    (\emph{Width}) The width of an $H$-decomposition is equal to $max\{|B_x|, x \in V(H)\} - 1$
\end{definition}
\begin{definition}
    (\emph{Treewidth}) The treewidth of a graph $G$ is the minimum width of a tree-decomposition of $G$. We denote it by $tw(G) = k$.
\end{definition}

\begin{definition}
    (\emph{Unicyclic graph}) A connected simple graph $G$ is unicyclic if it contains exactly one cycle.
\end{definition}

\begin{theorem}[Ramsey]
~\cite{ramsey} There exists a least positive integer $R(r, s)$ for which every assignment of $2$ colors to the edges of the complete graph on $R(r, s)$ vertices contains a blue clique on $r$ vertices or a red clique on $s$ vertices.
\label{ramsey_theorem}
\end{theorem}

\begin{theorem}[Erd\H{o}s-Szekeres]
    \cite{erdos} For given $s,r \in \mathbb{N}$, any sequence of distinct real numbers whose length is at least $sr + 1$ contains a monotonically increasing subsequence of length $s + 1$ or a monotonically decreasing subsequence of length $r + 1$.
    \label{erdos-szekeres}
\end{theorem}

\begin{theorem}
~\cite{hex} Let $H_n$ be the dual of an $n \times n$ hexagonal grid, where $V(H_n = \{1, ..., n\}^2$, and $uv \in E(H_n)$, where $u = [a, b] \in V(H_n)$ and $v = [c, d] \in V(H_n)$ if $|a - c| + |b - d| = 1$ or $|a - c| = |b - d| \in \{-1, 1\}$. Every assignment of $2$ colors to the vertices of $H_n$ contains a monochromatic path on n vertices.
\label{hex_lemma}
\end{theorem}

\begin{figure}[!ht]
    \centering
    \includegraphics[scale=.9]{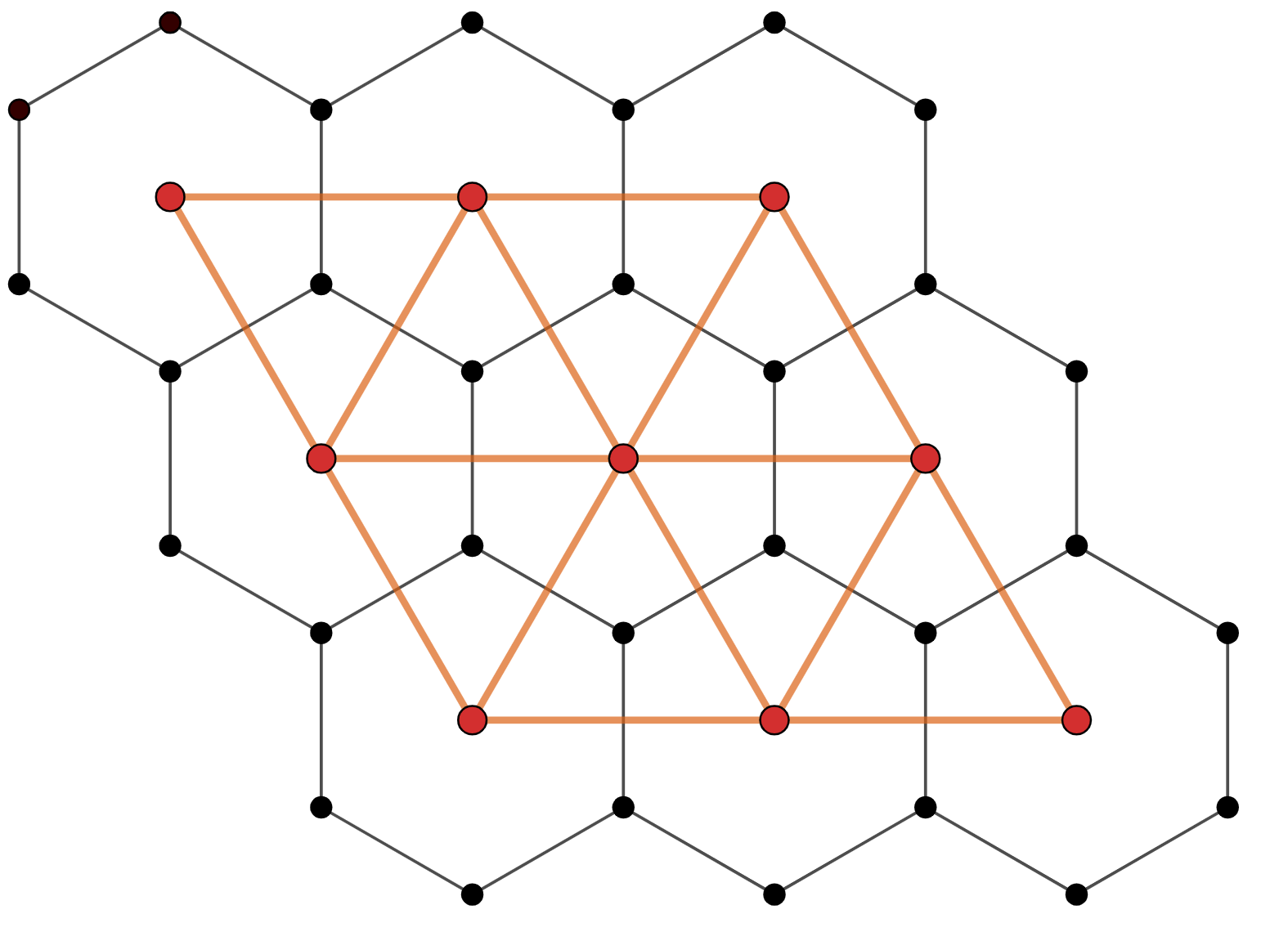}
    \caption{$3 \times 3$ hexagonal grid and $H_3$}
    \label{hexagonal_grid}
\end{figure}

\newpage
\section{Stack number}
\label{chapter_stack}
In this chapter, we give an overview of the stack number and notions related to it.
\begin{definition}
 (\emph{$k$-stack layout}) Let $G$ be a graph, $\prec$ a linear order on $V(G)$, and $\varphi$ a function $\varphi : E(G) \rightarrow \{1, ..., k\}$. We say that $(\prec, \varphi)$ is a $k$-stack layout of $G$ if no two edges $x$ and $y$ in $G$, such that $\varphi(x) = \varphi(y)$, are crossing in $\prec$.
\end{definition}
\begin{definition}                
 (\emph{Stack number}) Stack number of a graph $G$ is the minimal positive integer $k$, such that there exist a $k$-stack layout of $G$. We denote it by $sn(G) = k$.
\end{definition}

\begin{figure}[!ht]
\centering
    \input{pictures/stack_layout_example.tex}
\caption{$K_{2,3}$ and its possible $2$-stack layout}
\end{figure}
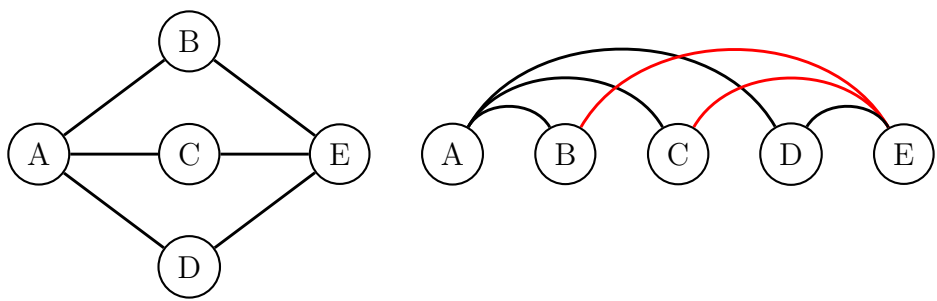 
Crossing edges are a problem for stack layouts. An easy observation shows, that if for a given ordering of $V(G)$, we have $k$ pairwise crossing edges, we need at least $k$ stacks to assign all the edges.
\begin{figure}[!ht]
\centering
    \input{pictures/twist.tex}
\caption{$3$ pairwise crossing edges}
\end{figure}
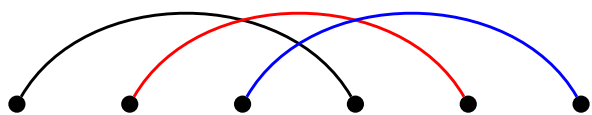 

With a different look at the stack number, we can see its association with the data structure stack.\\
Let $G$ be a graph and $(\prec, \varphi)$ its $k$-stack layout. We process the vertices of $G$ from left to right, as given by $\prec$, and process edges of $G$ using $k$ stacks. Assume we are currently processing vertex $a$. If there is an edge $ab$, where $a \prec b$, we insert this edge to its stack assigned by $\varphi$. If there is an edge $ca$, where $c \prec a$, we pop this edge from its stack. This means the edge $ca$ must be on the top of its stack. In a situation where we have multiple edges $ab_1$, $ab_2$, ..., $ab_n$ where $a \prec b_1 \prec b_2 \prec ... \prec b_n$. We put these edges into stacks as they are ordered by their right endpoints in order not to break the stack property.\\
Being able to process all the vertices and subsequently all the edges in $k$ stacks, whilst not breaking the stack property, is equivalent to forbidding crossed edges with the same assigned number by $\varphi$ in a $k$-stack layout $(\prec, \varphi)$.\\

\begin{figure}[!ht]
    \centering
    \includegraphics[scale=0.5]{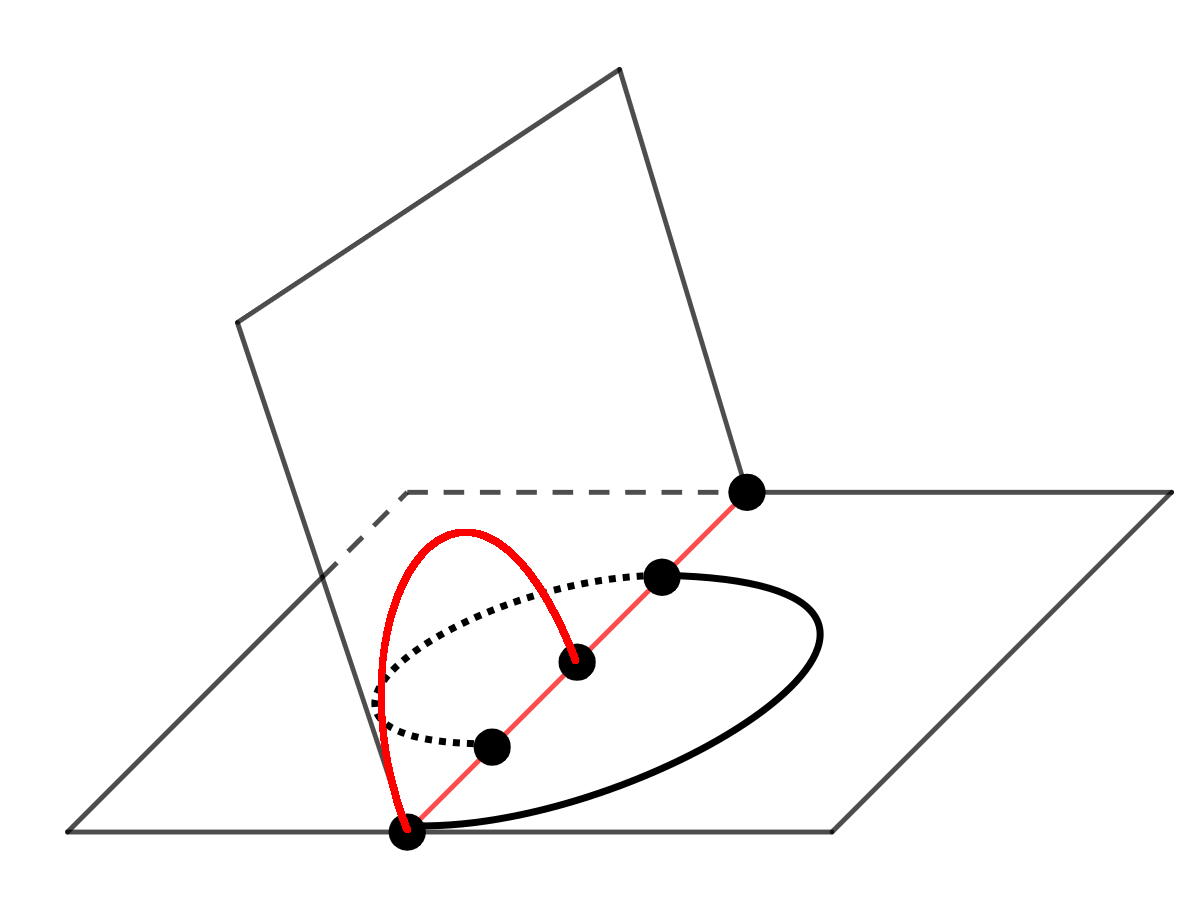}
    \caption{$3$-book embedding}
    \label{book_embedding}
 \end{figure}

We can also look at the stack number as the book embedding problem.
Embed the ordered vertices of a graph into a line called the back of the book. Edges are embedded into half-planes having one common boundary - the back of the book, as illustrated in Figure \ref{book_embedding}. These half-planes are called leaves or pages. There can be no crossing edges on a page. Our goal is to find the minimal number of pages necessary to embed a graph.

The book embedding, as it was defined in ~\cite{atneosen1968embeddability,  persinger1966subsets} by Atneosen and Persinger, is very similar to a stack layout. We can assign edges on a single page to the same stack and we have a valid layout. The only difference is, that we can embed edges into the back of the book itself. This way we do not need any pages to embed a path.

This was changed in the currently used definition by Kainen and Ollmann in ~\cite{kainen70sembedding, ollmann1973book}. This definition only allows edges on pages. With the current definition, the book thickness of any graph is equal to its stack number.
\subsection{Motivation}
Outside of the theory, stack number is important in many practical applications. For example in creating integrated circuits via VLSI design ~\cite{vlsi}. Other uses include direct interconnection networks~\cite{KAPOOR2002267}, fault-tolerant  processor arrays~\cite{rosenberg1983diogenes}, single row routing~\cite{so1974some} or sorting arrays using parallel stacks~\cite{even1971queues}.

\subsection{Stack number of graphs}
In this chapter, we look at the stack number of some classes of graphs and at classes of a given stack number.
\subsubsection{General graphs}
\begin{lemma}
    If there are $k$ pairwise crossing edges of $G$ in $\prec$, then $sn(G) \geq k$ for ordering $\prec$.
\end{lemma}
\begin{theorem}
~\cite{bernhart1979book} The stack number of a graph containing a cycle is equal to the maximal stack number of its $2$-connected components.
\end{theorem}
To find the stack number of a graph, it is sufficient to look at its $2$-connected components. However, it is NP-hard to find the stack number of a graph $G$ ~\cite{guan2022survey}. Even if we fix $k$ and the ordering of the vertices, it is NP-complete to decide whether $G$ accepts a $k$-stack layout with the given order ~\cite{vlsi}. If we draw the vertices in a circle as they are ordered, we can imagine assigning an edge to a stack, as coloring it a certain color. Now, this is the same as finding the chromatic number of circle graphs, which is NP-complete.

\begin{lemma}
    ~\cite{guan2022survey} Let $G$ be a simple graph and $V'$ a subset of $V(G)$ containing $k$ elements. Then $sn(G) \leq sn(G \setminus V') + k$.   
\label{lemma_general_sn}
\end{lemma}
\begin{proof}
    $G \setminus V'$ accepts $sn(G \setminus V')$-stack layout. A simple observation shows, that for any star $S$, $sn(S) = 1$. Since all the edges of $S$ share one vertex, they can never cross in any vertex ordering. Arbitrarily extend ordering of $G \setminus V'$ to a one of $G$. since edges adjacent to one vertex can be in the same stack, we only need at most $k$ new stacks. If we have an edge between vertices in $V'$, we can assign this edge to either of the two stacks.
\end{proof}
\begin{theorem}
    ~\cite{bernhart1979book} If $G$ is a simple graph, then $sn(G) \leq vc(G)$.
\end{theorem}
\begin{proof}
    This follows simply from Lemma \ref{lemma_general_sn}. If $V'$ is a vertex cover of $G$, then $G \setminus V'$ has no edges. Then clearly $sn(G) \leq vc(G)$.
\end{proof}

In ~\cite{bernhart1979book} Bernhart and Kainen showed that a $3$-stack graph can have an arbitrarily large genus. However, graphs of genus $g$ admit $O(g)$-stack layout. This result was published by Heath and Istrail in ~\cite{heath1984embedding}. This bound was lowered by Malitz in ~\cite{malitz1994genus} to $O(\sqrt{g})$. It can be shown, that the latter bound is tight.
\subsubsection{1-stack graphs}
\begin{theorem}
     ~\cite{bernhart1979book} If $G$ is a simple connected graph, then $sn(G) \leq 1$ iff $G$ is an outerplanar graph.
\end{theorem}
\begin{proof}    
    Assume a $1$-book embedding of $G$. It is a proper outerplanar drawing of $G$. Its planarity is given by the definition. If we extend the page to a full plane by adding another page, this added page is part of the outer face of $G$. Clearly, all the vertices are adjacent to this new page and therefore the outer face.

    Conversely, if we have an outerplanar graph $G$, we can order its vertices as they appear on the outer face (every vertex only once). This forms a proper ordering for a $1$-stack layout of $G$.\\
\end{proof}
    
    To recognize a $1$-stack graph we simply need to check for outerplanarity. This can be done in linear time~\cite{syslo1979characterizations}.   
\subsubsection{2-stack graphs}
\begin{theorem}
   ~\cite{bernhart1979book} If $G$ is a simple connected graph, then $sn(G) \leq 2$ iff $G$ is subhamiltonian.
   \label{2-stack_theorem}
\end{theorem}
\begin{proof}
    
   $2$-stack graphs can be embedded in $2$ pages. The union of two pages forms a topological space equivalent to the plane. By the definition of book embedding, the drawing preserves planarity. This means every $2$-stack graph is planar. Moreover, we can augment the graph to a planar Hamiltonian graph. This can be achieved by connecting any two vertices laid out next to each other which were previously not connected and adding an edge between the first and the last vertex, if it was not already there. Adding these edges clearly preserves planarity and forms a Hamiltonian cycle.

   Conversely, let $G$ be a subhamiltonian graph. By the definition, $G$ can be edge augmented to create a Hamiltonian graph while preserving planarity. Order the vertices along the back of the book as they appear on the Hamiltonian cycle. Because of the planarity both edges on the inside and outside of the Hamiltonian cycle form an outerplanar graph. We can draw the inside edges on one page and the outside edges on another. This gives a valid $2$-page embedding.\\
\end{proof}
   
   Recongnizing subhamiltonian graphs is NP-complete~\cite{wigderson1982complexity}. Therefore, recognizing $2$-stack graphs is NP-complete as well.

\subsubsection{Planar graphs}
The stack number of planar graphs is a very well-researched topic. From the $2$-stack graphs, we know that some planar graphs have stack number 2. There are however maximal planar graphs (triangulations) which do not have a Hamiltonian cycle. For example the Goldner-Harary graph. This means there are planar graphs that require at least a $3$-stack layout.\\
First, it was conjectured by Bernhart and Kainen\cite{bernhart1979book} that the stack number of planar graphs is unbounded. The first upper bound on the stack number of planar graphs was $O(\sqrt{n})$ where $n$ is the number of vertices. The upper bound of $9$ was reached by Buss and Shor in ~\cite{buss1984pagenumber}. This bound was lowered to 7 by Heath in ~\cite{heath1984embedding}. Finally, Yannakakis proved the final bound.
\begin{theorem}[Yannakakis]
~\cite{yannakakis1986four, yannakakis2020planar} If $G$ is a planar graph, then $sn(G) \leq 4$. Moreover, there are planar graphs with stack number $4$.
\end{theorem}
\begin{sketch_proof}
    
The $4$-stack algorithm was inspired by Heath's algorithm. They both work in a similar way, peeling the layers of the graph.\\
The algorithm's input is a quasi-triangulation $G$, with some requirements for the structure of the outer face. We partition the vertices of $G$ into levels. Level $0$ are vertices on the outer face. In level $1$ are the vertices adjacent to level $0$ and so on. This way, we sort the vertices of $G$ into their according levels. The algorithm then recursively "peels" the graph by levels and constructs the $4$-stack layout level by level.

The structure of a $4$-stack planar graph is rather complicated. Yannakakis provides in his paper this intuition. Let $Q$ be an outerplanar graph with $4$ vertices on the outer face. We then attach a copy of $Q$ to every edge of a triangulation $T$, where $sn(T) = 3$. Intuition on why $3$ stacks no longer suffice is as follows. Let's say $Q$ has a property, that some of its vertices have to be laid out in a place, right next to the two vertices of the edge it is attached to. Then we have $2n$ places, but $3n-6$ edges, where $n$ is the number of vertices. This means, that for large enough $n$, we can not lay out all the copies of $Q$. 
\end{sketch_proof}

\subsubsection{Complete graphs}
\begin{theorem}
    \cite{vlsi} If $G$ is a complete graph, then $sn(G) = \lfloor |V(G)|/2 \rfloor$
\end{theorem}
\begin{proof}
    Fix any ordering $\sigma = v_1, v_2, ..., v_n$ of $V(G)$. Edges $\{i, \lfloor n/2 \rfloor + i\}$ for $1 \leq i \leq \lfloor n/2 \rfloor$ are pairwise crossing. This means $sn(G) \geq \lfloor n/2 \rfloor$.

    Layout vertices of $K_n$ on a circle as they are ordered. For any vertex $w$, call $w_i$ the ith vertex in the clockwise direction and $w_{-i}$ the ith vertex in the anti-clockwise direction from $w$. For each each vertex $w$, such that $v_1 \preceq w \preceq v_{\lfloor n/2 \rfloor}$ define a path $P_w = w, w_1, w_{-1}, ..., w_{\lfloor n/2 \rfloor - 1},$ $w_{-(\lfloor n/2 \rfloor - 1)}, w_{\lfloor n/2 \rfloor}$, as illustrated in Figure \ref{paths_in_complete}.
    \begin{figure}[!ht]
    \centering    
    \input{pictures/circular_layout.tex}
    \caption{Circular layout of $G$ with two paths}
    \label{paths_in_complete}
\end{figure}
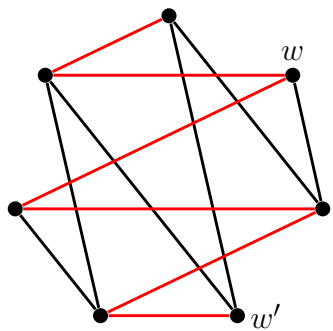
    If we cut the circle between any two vertices and spread it into a line, we get a $1$-stack layout of $P_w$. If there were crossing edges in such ordering, there would be crossing edges in the circle embedding as well. This is not possible with our construction of $P_w$.

    Every vertex $w$ is an endpoint of exactly one path. In every other path, valence of $w$ is $2$. This means a graph containing all the paths has $n-1$ edges coming from every vertex $w$. Also no two paths share an edge. This means every edge of $G$ is in exactly one path. We can assign edges of one path to a single stack and spread the circle into a line as described above to obtain a $\lfloor n/2 \rfloor$-stack embedding of $G$.
\end{proof}
\subsubsection{Trees}
\begin{theorem}
    ~\cite{vlsi} If $G$ is a tree, then $sn(G) = 1$.
\end{theorem}
\begin{proof}
    Root $G$ in an arbitrary vertex $r$. Scan $G$ in a depth-first manner from $r$, and lay out vertices, as they are encountered. Simple observation shows, that no two edges can be crossing in this order. This means we have a valid $1$-stack layout.
\end{proof}
\subsubsection{X-trees}
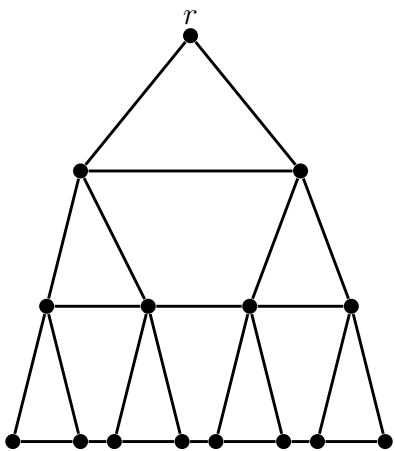
\begin{figure}[!ht]
    \centering
    \input{pictures/xtree.tex}
    \caption{Depth-$3$ $X$-tree}
    \label{x-tree}
\end{figure}

Let $G$ be a complete binary tree of depth $d$ rooted in a vertex $r$. Depth-$d$ $X$-tree $X(d)$ is an edge augmentation of $G$ created the following way. In every layer of $G$ connect every vertex to the closest vertex to the right in the same layer. See Figure \ref{x-tree}.
\begin{theorem}
    ~\cite{vlsi} If $G$ is a depth-$d$ X-tree for $d > 2$, then $sn(G) = 2$.
\end{theorem}
\begin{proof}
    Add two edges from left-most and right-most vertices in the last layer to the root $r$. We can easily find a Hammiltonian cycle by traversing the paths within layers from the top to the bottom in alternating order and returning by one of the edges. This means $G$ is subhamiltonian. By Theorem \ref{2-stack_theorem} $G$ accepts a $2$-stack layout.
    
    One can see, that for $d > 2$, $X(d)$ is no longer outerplanar and therefore does not accept $1$-stack layouts.
\end{proof}
\subsubsection{k-trees}
Every complete graph $K_k$ is a $k$-tree. Adding a new vertex to a $k$-tree that is adjacent to $k$ vertices forming a $K_k$ results in a $k$-tree. $k$-trees are maximal graphs of treewidth $k$.

\begin{theorem}
    ~\cite{ganley2001pagenumber} If $G$ is a $k$-tree, then $sn(G) \leq k + 1$.
\end{theorem}
\begin{proof}
    A $G$ has a tree decomposition of width $k$, with an underlying tree $T = (I, F)$. Order the vertices of $G$ by a depth-first search of $T$ from an arbitrary vertex. The ordering of $v$ is determined by the first time $i \in I$ with $v \in X_i$ is encountered (vertices with the same discovery time are ordered arbitrarily). By the properties of tree decomposition, every $v \in G$ induces a non-empty subtree $T_v$ of $T$, such that $i \in T_v$ iff $v \in X_i$. By ~\cite{gavril1974intersection} the intersection graph of subtrees of a tree is a chordal graph. Chordal graphs are perfect, which means their chromatic number is equal to their maximum clique size. Since $G$ is a $k$-tree, there is no clique greater than $k + 1$, hence we can color the subtrees (vertices of $G$) using $k + 1$ colors. Now we color the edges in the following way. Assign $uv \in E$ the color of $u$ if $u \prec v$, assign it the color of $v$ otherwise.
    Now no two edges of the same color cross. Suppose for contradiction $ab$ and $cd$ have the same color and WLOG $a \prec c \prec b \prec d$. This means $T_a$ and $T_c$ intersect, which is  contradiction with the assumption that $ab$ and $cd$ have the same color. The presented ordering and coloring yields a proper $k + 1$-stack layout of $G$.
\end{proof}
\begin{theorem}
    ~\cite{muder1988pagenumber} There are $k$-trees that require $k$ stacks.
\end{theorem}
\begin{sketch_proof}
Such graphs have the following structure. Start with a complete graph $K_k$ and add $mk$ vertices adjacent to the original $K_k$. We denote these graphs $K_{mk,k}$. By ~\cite{muder1988pagenumber} $sn(K_{mk,k}) \geq (\frac{m}{m + 1})n$. Clearly as $m$ approaches infinity, the lower bound for stack number of $G$  approaches $n$.
\end{sketch_proof}

\newpage
\section{Queue number}
\label{chapter_queue}
In this chapter, we look at the invariant queue number. We give its formal definition and define related notions.
\begin{definition}
 (\emph{$k$-queue layout}) Let $G$ be a graph, $\prec$ a linear order on $V(G)$, and $\varphi$ a function $\varphi : E(G) \rightarrow \{1, ..., k\}$. We say that $(\prec, \varphi)$ is a $k$-queue layout if no two edges $x$ and $y$, such that $\varphi(x) = \varphi(y)$ are nested in $\prec$.
\end{definition}
\begin{definition}                
 (\emph{Queue number}) The queue number of a graph $G$ is the minimal positive integer $k$, such that there exist a $k$-queue layout of $G$. We denote it by $qn(G) = k$.
\end{definition}
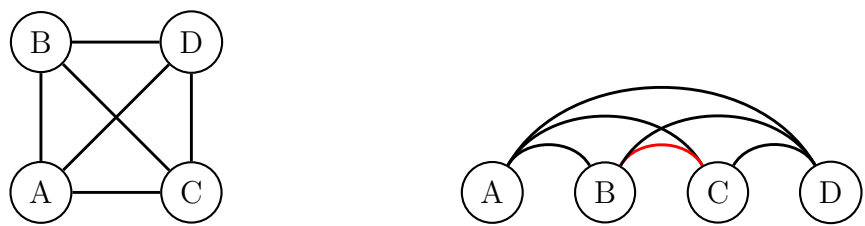
\begin{figure}[!ht]
    \centering
    \input{pictures/queue_example}
    \caption{$K_4$ and its possible $2$-queue layout}
\end{figure} 
Same as with the stack number, we can look at this problem as assigning edges to queues while scanning the ordered vertices. Again, not breaking the queue property is equivalent to not having nested edges with the same assigned number by $\varphi$ in a $k$-queue layout $(\prec, \varphi)$.\\

Similarly, as $k$-twist forms an obstacle for stack layouts, pairwise nested edges are a problem for queue layouts. We call a set of $k$ pairwise strictly nested edges a $k$-rainbow.
\begin{figure}[!ht]
\centering
\trimbox{0pt, 5pt, 0pt, 10pt}{
    \input{pictures/rainbow}
}
\caption{$3$-rainbow}
\end{figure}
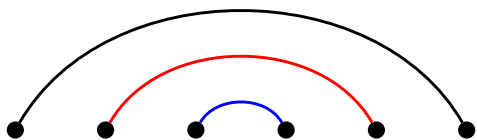 

It is NP-complete to determine the queue number of a graph ~\cite{ganian4}. Determining the queue number of a graph for fixed ordering is easier than determining its stack number.
\begin{theorem}
    ~\cite{laying_queues} Let $G$ be a simple graph and $\sigma$ a fixed ordering of $V(G)$. The largest rainbow size of $\sigma$ is equal to the queue number of $G$ for $\sigma$.
\end{theorem}
Heath and Rosenberg also provide an algorithm, which finds a $k$-queue layout of $G$ for ordering $\sigma$ in $O(m\ log(log(n)))$ time, where $k$ is the size of the larges rainbow in $\sigma$, $m$ is the number of edges, and $n$ the number of vertices of $G$.
\subsection{Motivation}

The motivation is very similar to that of the stack number. Many of the problems solved using stacks can be solved with different approach using queues. Some of the problems include the fault-tolerant processor arrays ~\cite{rosenberg1983diogenes} or permuting objects using queues ~\cite{tarjan1972sorting}.
\subsection{Queue number of graphs}
In this section we look at the class of $1$-queue graphs and the queue number of some classes of graphs.
\subsubsection{1-queue graphs}
\begin{definition}
    (\emph{Leveled planar graph}) Consider a normal Cartesian coordinate system $(x, y)$. For $i \in \mathbb{N}$ let $l_i$ be a vertical line $l_i=\{(i, y)|y \in \mathbb{R}\}$. A simple graph $G$ is a leveled planar graph, if $V(G)$ can be partitioned into sets $V_1, V_2, ..., V_n$ and $G$ can be embedded in the plane, such that all vertices of $V_i$ are on the line $l_i$ and for every edge $ab \in E(G)$ it holds that if $a \in V_i$, then $b \in V_{i+1}$ or $b \in V_{i-1}$.
\end{definition}
Leveled planar graphs have an induced order. We scan the lines $l_i$ from left to right each line from bottom to top ordering the vertices as they are encountered. We call $v_{b_i}$ the first vertex encountered on line $l_i$ and $v_{t_i}$ the last vertex on line $l_i$. If there are edges between parts $V_i$ and $V_{i+1}$,  we call $v_{s_i}$ the first vertex adjacent to $V_{i+1}$. Otherwise $v_{s_i} = v_{t_i}$.
\begin{definition}
    (\emph{Level $i$ arch}) A level $i$ arch for a leveled planar graph $G$ is an edge between $v_{t_i}$ and $v_j$, where $b_i \leq j \leq min\{t_i - 1, s_i\}$, where $t_i$ is the index of the last (topmost) vertex encountered on line $l_i$ and $s_i$ is the index of the first vertex on line $l_i$ adjacent to some vertex on line $l_{i+1}$. If there is no edge between lines $l_i$ and $l_{i+1}$, then $s_i = t_i$.
\end{definition}
\begin{definition}
    (\emph{Arched leveled planar graph}) A leveled planar graph augmented by $0$ or more arches is called an arched leveled planar graph. Adding arches to a leveled planar graphs preserves planarity.
\end{definition}
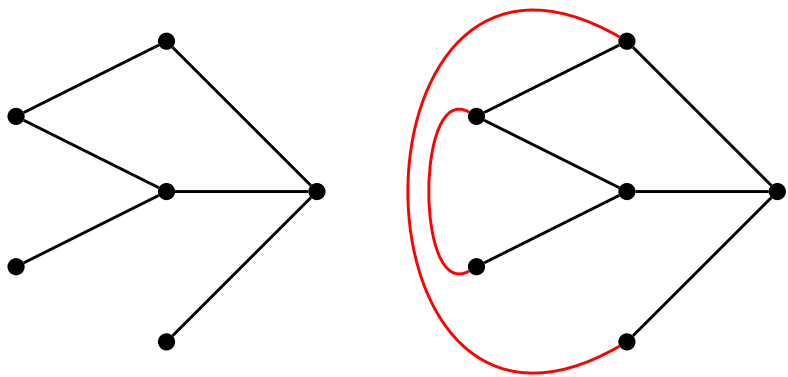
\begin{figure}[!ht]
\begin{center}
\trimbox{60pt 55pt 0pt 60pt}{
\input{pictures/leveled}
\hfill
\input{pictures/arched_leveled}
}
\end{center}
\caption{A leveled planar and an arched leveled planar graph}
\end{figure} 
In ~\cite{laying_queues} Heath and Rosenberg proved that $1$-queue graphs are exactly arched leveled planar graphs.

\begin{lemma}
    ~\cite{laying_queues} Every leveled planar graph is a $1$-queue graph.
    \label{lpg_lemma}
\end{lemma}
\begin{proof}
    The order induced the on leveled planar graph $G$ yields a $1$-queue layout of $G$. This is equivalent to saying that there are no nested edges in this ordering. Pick any two edges on four distinct vertices $s_1r_1$ and $s_2r_2$. Assume they are nested. Clearly $s_1$ is on the same line as $s_2$ and both $r_1$ and $r_2$ are on the same neighboring line. This means $s_1r_1$ and $s_2r_2$ are crossing in the leveled planar embedding, which is a contradiction. 
\end{proof}
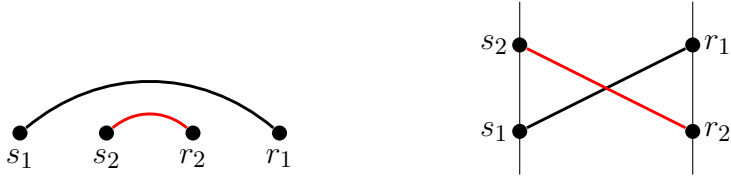
\begin{figure}[!ht]
    \centering
    \hfill
    \subfigure{\input{pictures/nested_leveled}}
    \hfill
    \subfigure{\input{pictures/crossed_leveled}}
    \hfill
    \hfill
    \caption{Nested edges in leveled planar embedding}
    \label{nested_planar}
\end{figure}
    
\begin{lemma}
    ~\cite{laying_queues} Every arched leveled planar graph is a $1$-queue graph.
    \label{alpg}
\end{lemma}
\begin{proof}
    By Lemma \ref{lpg_lemma}, it is enough to show, that arches are not nested with any other edge. No two arches can nest with each other. If the arches are on the same line, they share a vertex and therefore can not nest. If they are on different lines, they are separated with respect to $\prec$, and can not be nested. Since arches are contained within one line and leveled edges are between two neighboring lines, they can not nest as well. Hence, we have a $1$-queue layout.
\end{proof}

\begin{lemma}
    ~\cite{laying_queues} Every $1$-queue graph is an arched leveled planar graph.
    \label{alpg_back}
\end{lemma}
\begin{proof}
    WLOG $G$ is a connected $1$-queue graph. Let $\sigma = v_1, v_2, ..., v_n$ be ordering of its $1$-queue layout. Partition $V(G)$ into $V_1, ... V_m$ as follows. $V_1 = \{v_1\}$ and $v_{s_1} = v_{t_1} = v_1$. For $i > 1$ let $V_i = \{v_{b_i}, ..., v_{t_i}\}$, where $v_{b_i} = v_{t_{i-1} + 1}$ and $v_{t_i}$ is the last vertex incident to a vertex in $V_{i-1}$. Label $v_{s_i}$ the leftmost vertex adjacent to some vertex to the right of $t_i$. By construction, we only have edges between vertices in neighboring parts (set $E_l$) or vertices within one part (set $E_a$).
    
    The graph $(V(G), E_l)$ has a leveled planar embedding. We put vertices in one part on a single line from bottom to top. Edges in $E_l$ can not cross, as they would need to be nested similarly to Figure \ref{nested_planar}.

    Edges in $E_a$ are all arches. By construction, they are all within one part. Let $sr$ be an edge of $E_a$, where $s \prec r$ and $s, r \in V_j$. Observation shows that $s \prec min\{v_{t_j-1}, s_j\}$. Should $s$ be on the right of any of those vertices, it would form a $2$-rainbow with another edge. By construction, $v_{t_j}$ is connected to a vertex in $V_{j-1}$, therefore $r = t_j$ as otherwise these two edges would nest. This means all edges in $E_a$ are arches, thus we have a valid arched leveled planar embedding of any $1$-queue graph $G$.\\
\end{proof}

By Lemmas \ref{alpg} and \ref{alpg_back} we get the final result.
\begin{theorem}
    ~\cite{laying_queues} $G$ is a $1$-queue graph iff $G$ is an arched leveled planar graph.
\end{theorem}

To recognize a $1$-queue graph $G$, we need to check, if $G$ accepts an arched leveled planar embedding. In ~\cite{laying_queues} Heath and Rosenberg show, that this problem is NP-complete. 
\subsubsection{Complete graphs}
\begin{theorem}
    ~\cite{laying_queues} If $G$ is a complete graph, then $qn(G) = \lfloor n/2 \rfloor$
\end{theorem}
\begin{proof}
    Fix any ordering $\sigma = 1, 2, ..., 3$ of $K_n$. It contains $\lfloor n/2 \rfloor$-rainbow on edges $\{i, n+1-i\}$ for $1 \leq 1 \leq \lfloor n/2 \rfloor$, therefore $qn(K_n) \geq \lfloor n/2 \rfloor$.

    The length of an edge $ij$ is $|i - j|$. There are all lengths of edges from $1$ to $n - 1$. Simple observation shows, that edges of the same size or edges whose lengths differ by one can not be nested. This means for $i \in \{1, ..., \lfloor n/2 \rfloor \}$ we can assign edges of length $2i - 1$ and $2i$ to the same queue, creating $\lfloor n/2 \rfloor$-queue layout.
\end{proof}

\subsubsection{Complete bipartite graphs}
\begin{theorem}
    ~\cite{laying_queues} For all $m, n \in \mathbb{N}$, the queue number of the complete bipartite graph $K_{m, n}$ is $ qn(K_{m, n}) = min\{\lceil m/2 \rceil, \lceil n/2 \rceil \}$
\end{theorem}
\begin{proof}
    
    Assume $K_{m,n}$ has vertex set partitioned into two parts $A=\{a_1, ..., a_m\}$ and $B = \{b_1, ..., b_n\}$, where WLOG $m \leq n$.

    First, we show that $qn(K_{m, n}) \leq \lceil m/2 \rceil$. Fix and ordering $\sigma = a_1, ..., a_{\lceil m/2 \rceil},\\ b_1, ..., b_n, a_{\lceil m/2 \rceil + 1}, ..., a_m$. For $i \in \{1, ... \lceil m/2 \rceil\}$ the $i$th queue contains edges $a_ib_j$ and $\{a_{m+1-i}, b_j\}$ for $1 \leq j \leq n$. This forms a valid $\lceil m/2 \rceil$-queue layout.

    Now we prove the lower bound. Let $\sigma$ be any ordering of $K_{m,n}$. Assume vertices of $A$ appear in order $a_1, ..., a_m$ and vertices of $B$ appear in order $b_n, ..., b_1$. Since rotating a layout does not change anything, we may assume $a_{\lceil m/2 \rceil} \prec b_{\lceil n/2 \rceil}$. This means edges $\{a_ib_i | 1 \leq i \leq \lceil m/2 \rceil \}$ form  $\lceil m/2 \rceil$-rainbow, therefore $qn(K_{m, n} \geq \lceil m/2 \rceil)$
\end{proof}
\begin{figure}[!ht]
    \centering
    \input{pictures/k44layout}
    \caption{$q_1$ in the $2$-queue layout of $K_{4,4}$}
\end{figure}
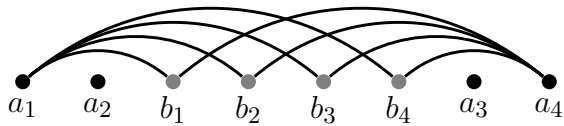

\subsubsection{Planar graphs}
In ~\cite{laying_queues} Heath and Rosenberg conjectured, that planar graphs have bounded queue number. The first bound $O(\sqrt{m})$ was reached in \cite{comparing}. This bound was lowered to $log^2(V(G))$ in ~\cite{di2013queue}. Finally, the conjecture was proved correct by Dujmovi\'c $et$ $al.$ in ~\cite{dujmovic2020planar}.
\begin{theorem}
    ~\cite{dujmovic2020planar} If $G$ is a planar graph, then $qn(G) \leq 49$.
    \label{theorem_planar_qn}
\end{theorem}
The proof uses a layered partition of $G$. It consists of partition $\mathcal{P}$ of $V(G)$, such that there are edges only between neighboring parts or edges contained within one part and a layering of $V(G)$, such that the number of the same layer vertices in one part is bounded. The quotient graph $G/ \mathcal{P}$ in the proof also has bounded treewidth. This is important because of the following result.
\begin{theorem}
    ~\cite{wiechert2016queue} If $G$ has treewidth $k$, then $qn(G) \leq 2^k - 1$.
\end{theorem}
The bound in Theorem \ref{theorem_planar_qn} was lowered to $42$ in ~\cite{bekos2021queue} using the same tools as the previous bound.
\subsubsection{Trees}
\begin{theorem}
    ~\cite{laying_queues} If $G$ is a tree, then $qn(G) \leq 1$.
    \label{qn_trees}
\end{theorem}
\begin{proof}
    Root $G$ at arbitrary vertex $r$. If we lay $G$ out in a breadth-first manner, no edges can be nested. This yields a proper $1$-queue layout. \end{proof}
The BFS layering of $G$ also yields a proper leveled planar embedding with only the root $r$ in the first line $l_1$.
\label{trees}
\subsubsection{X-trees}
\begin{theorem}
    ~\cite{laying_queues} If $G$ is a depth-$d$ $X$-tree for $d > 2$, then $qn(G) = 2$.
\end{theorem}
\begin{proof}
    Scan $G$ starting in the root $r$ in breadth-first manner. Visit the children of a vertex from left to right, laying out the vertices as they are encountered. The underlying complete binary tree is laid out in breadth-first manner, therefore we can assign its edges to a single queue, as in section \ref{trees}. The edges of paths within layers can be assigned to the second queue.

    To prove the lower bound we show, that $qn(X(2)) > 1$. For any edge $xy$ in $G$ define operation $hat(xy)$ as adding a vertex $z$ called peak and edges $xz$ and $yz$ to $G$.

    Start with a cycle $C_4 = u_1, u_2, u_3, u_4, u_1$. Hat any three edges of $C_4$. The resulting graph is isomorphic to $X(2)$. Suppose $X(2)$ has a $1$-queue layout with ordering $\prec$. WLOG $u_1$ is the leftmost vertex of $C_4$ in $\prec$. Only $u_3$ can be the rightmost vertex in $\prec$, as otherwise edges would nest. We can assume $u_1 \prec u_2 \prec u_4 \prec u_3$. Either $u_1$ or $u_3$ has both incident edges hatted. We may assume $u_1u_2$ and $u_1u_4$ are hatted. Let $p$ be the peak of $u_1u_4$. $p$ can only be place between $u_1$ and $u_2$ without creating nested edges. However, now we can not place the peak of $u_1u_2$ anywhere without creating nested edges. This is a contradiction with the assumption, that $X(2)$ has a $1$-queue layout. Since every graph $X(d)$ for $d \geq 2$ has $X(2)$ as a subgraph, $qn(X(d)) > 1$.
\end{proof}

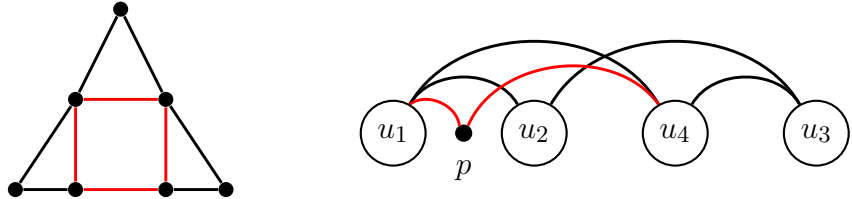
\begin{figure}[!ht]
    \centering
    \hfill
    \subfigure{\input{pictures/c4a}}
    \hfill
    \subfigure{\input{pictures/c4b}}
    \hfill
    \hfill
    \caption{$C_4$ with three hatted edges and $C_4$ laid out with the peak $p$ of $(u_1, u_4)$}
\end{figure}
\subsubsection{Unicyclic graphs}

\begin{theorem}
    ~\cite{laying_queues} If graph $G$ is unicyclic, then $qn(G) = 1$.
\end{theorem}
\begin{proof}
    $G$ contains a cycle $C = u_1, u_2, ..., u_k, u_1$. If $k$ is even, partition $C$ into $k/2 + 1$ parts $U_1 = \{u_1\}$, $U_i = \{u_i, u_{k-i+2}\}$, $U_{(k/2)+1} = \{u_{(k/2)+1}\}$, for $1 < i <(k/2)+1$. If $k$ is odd partition $C$ into $(k+1)/2$ parts $U_i = \{u_i, u_{k-i+1}\}$, $U_{(k+1)/1} = \{u_{(k+1)/2}\}$ for $1 \leq i < (k+1)/2$. Both cases yield an arched leveled planar embedding. The only arch is $u_1u_k$ for odd $k$.
\end{proof}

\begin{figure}[!ht]
    \centering
\trimbox{0pt 0pt 0pt 5pt}{
    \input{pictures/arched_cycle_a}
    \hfill
    \hfill
    \input{pictures/arched_cycle_b}
}
\caption{Arched leveled planar embedding of $C_7$ and $C_6$}
\end{figure}
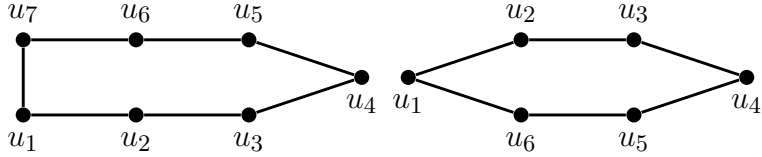
Define $G'$ a subgraph of $G$ without edges of $C$. $G'$ contains a tree $T_i$ for every vertex $u_i \in C$. Root every tree $T_i$ at $u_i$. We can convert the arched leveled planar embedding of $C$ to one of $G$ by expanding $u_i$ to $T_i$ in breadth-first manner, as in subsubsection \ref{trees}. 
\subsubsection{Cartesian products}
\begin{definition}
    (\emph{Strict queue layout}) A queue layout is strict if there are no two edges $ab$ and $ac$ in the same queue, such that $a \prec b \prec c$ or $ c \prec b \prec a$. The strict queue number is defined analogously to the queue number. We denote it by $sqn(G)$.
\end{definition}
\begin{figure}[!ht]
    \centering
    \hfill
    \begin{tikzpicture}[every edge/.append style={line width = 1.1pt}, every node/.style={circle,draw}]
        \node (1) at (0,0) {$a$};
        \node (2) at (2,0) {$b$};
        \node (3) at (4,0) {$c$};
        \path (1) edge[bend left=60, color=red] (2);
        \path (1) edge[bend left=60](3);
    \end{tikzpicture}
    \hfill
    \begin{tikzpicture}[every edge/.append style={line width = 1.1pt}, every node/.style={circle,draw}]
        \node (1) at (0,0) {$c$};
        \node (2) at (2,0) {$b$};
        \node (3) at (4,0) {$a$};
        \path (1) edge[bend left=60] (3);
        \path (2) edge[bend left=60, color=red](3);
    \end{tikzpicture}
    \hfill
    \hfill
    \caption{$2$-strict queue layout}
\end{figure}
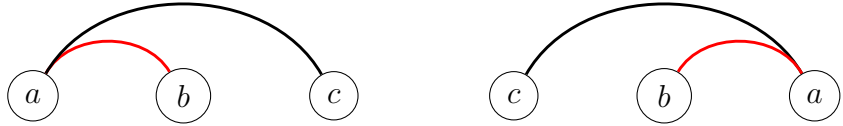
\begin{theorem}
    ~\cite{wood2005queue} If $G$ and $H$ are simple graphs, then $qn(G \square H) \leq sqn(G) + qn(H)$.
\label{wood_product}
\end{theorem}
\begin{proof}
    Let $\prec_s$ be the ordering in a $sqn(G)$-strict queue layout of $G$ and $\prec_q$ the ordering in a $qn(H)$-queue layout of $H$. Construct the vertex ordering $\prec$ of $X := G \square H$ the following way. $(v, a) \prec (w, b)$ iff $v \prec_s w$ or $v = w$ and $a \prec_q b$.
    
    The copies of $H$ in $X$ are separated with respect to $\prec$ and their relative order is equal to $\prec_q$. This means we can put edges of every copy of $H$ in $X$ to the same $qn(H)$ queues.

    Suppose two distinct $G$-edges $(a, u)(b, u)$ and $(c, v)(d, v)$, where $ab$ and $cd$ are in the same strict queue in $sqn(G)$-queue layout of $G$. Since the edges are in a strict queue and can not share a vertex we may assume $a \prec_s c$ and $b \prec_s d$, thus $(a, u) \prec (c, v)$ and $(b, u) \prec (d, v)$. Hence, for every strict queue in $G$, the corresponding $G$-edges of $X$ form a strict queue in $\prec$.
    
    \begin{figure}[!ht]
        \centering
        \input{pictures/cartesian}
        \caption{Ordering $\prec$ of $S_3 \square P_3$ with black $S_3$-edges and red $P_3$-edges}
    \end{figure}
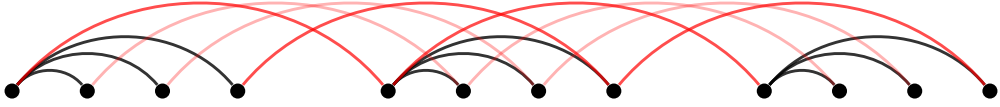
\end{proof}
\newpage
\section{A stack number bound on the queue number}
\label{chapter_oposite}
It is still an open problem whether the queue number is bounded by the stack number of the same graph. This problem was researched in ~\cite{comparing, dujmovic2005stacks, dujmovic2020planar}.
The first result was reached by Heith, Leighton, and Rosenberg.
\begin{theorem}
    ~\cite{comparing} Every $1$-stack graph accepts a $2$-queue layout.
\end{theorem}
\begin{sketch_proof}
    Let $G$ be a maximal $1$-stack graph, and therefore an outerplanar graph. Since $G$ is maximal, we may assume it has a unique outerplanar embedding with all its vertices on the outer face and the boundary of the face is the unique Hammiltonian cycle of $G$. Ordering is given by this cycle. Vertices are then assigned to layers by BFS. It can be shown that intra-layer edges are not nested with each other and therefore can be in a single stack. The same holds for inter-layer edges, thus obtaining a $2$-stack layout.
\end{sketch_proof}

\begin{theorem}
    ~\cite{dujmovic2020planar} $2$-stack graphs have bounded queue number,
\end{theorem}
\begin{proof}
    Dujmovic $et\ al.$ proved that planar graphs have bounded queue number. Since $2$-stack graphs are subhamiltonian and therefore planar, it follows that $2$-stack graphs have bounded queue number as well.
\end{proof}

The queue number of general graphs was studied by Dujmovi\'c and Wood in ~\cite{dujmovic2005stacks}.
\begin{theorem}
    ~\cite{dujmovic2005stacks} Every graph $G$ has a $3$-stack subdivision.
    \label{sub_theorem}
\end{theorem}
\begin{sketch_proof}
    We show the structure of a subdivision of a simple graph $G$ which accepts a $3$-stack layout. Start with a $2$-subdivision $G'$ of $G$. Let $\prec$ be any linear ordering of $V(G)$. Extend $\prec$ to $\prec'$ by ordering vertices adjacent to $v \in V(G)$ immediately right to $v$. Clearly, the edges between the original vertices and the vertices adjacent to them can not cross in $\prec'$ and can all be in the same stack. The remaining edges form a matching. Assigning them to two stacks is the same as drawing the edges in a plane without crossing such that the vertices are fixed on a line $l$. This is possible, every time and edge has to  cross the line $l$, we simply subdivide the edge. For example assume we have vertices $a \prec b \prec c \prec d \prec e$ and edges $be$, $ce$ adn $ad$. Clearly $be$ and $ce$ are in different pages. The edge $ad$ can not be embedded in either page, as it would cross either $be$ or $ce$ in $\prec$. To embed $ad$, it needs to cross the line between the pages, which can be achieved by subdividing $ad$.
\end{sketch_proof}

\begin{lemma}
    ~\cite{dujmovic2005stacks} Let $D$ be a $q$-queue subdivision of $G$ with at most $k$ division vertices per edge. Then $qn(G) \leq \frac{1}{2}(2q + 2)^{2k} - 1$.
    \label{sub_lemma}
\end{lemma}
\begin{theorem}
    ~\cite{dujmovic2005stacks} The queue number of a graph $G$ is bounded by its stack number iff bipartite $3$-stack  graphs have bounded queue number.
\end{theorem}

\begin{sketch_proof}
    If the queue number is bounded by the stack number, then it immediately follows that bipartite $3$-stack graphs have bounded queue number. To prove the other implication, assume bipartite $3$-stack graphs have bounded queue number. By Theorem \ref{sub_theorem}, every graph has a $3$-sack subdivision. We can construct such subdivision with an odd amount of division vertices per edge. This means such subdivision would be bipartite and therefore would have bounded queue number. With a careful construction, the maximal number of division vertices per edge is logarithmic in $sn(G)$. By Lemma \ref{sub_lemma} $qn(G)$ is bounded by a polynomial function of $sn(G)$.
\end{sketch_proof}

\begin{figure}
    \centering
    \trimbox{3pt, 1pt, 3pt, 1pt}{
    \includegraphics[scale=1.3]{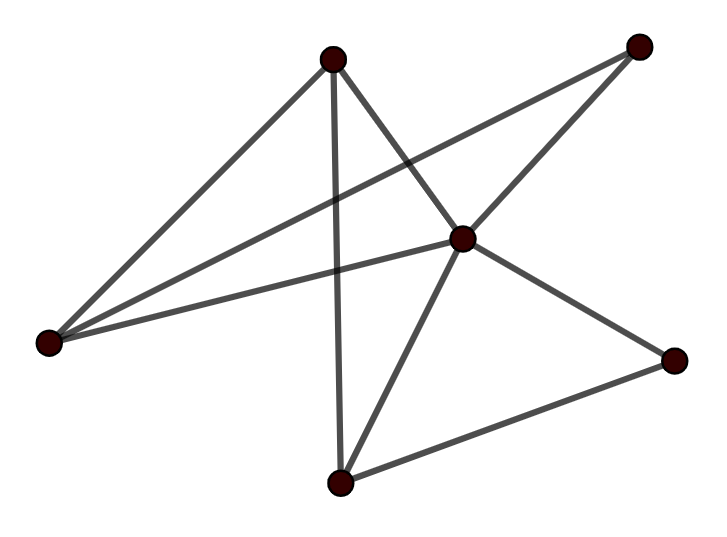}
    }
    \includegraphics[scale=1.3]{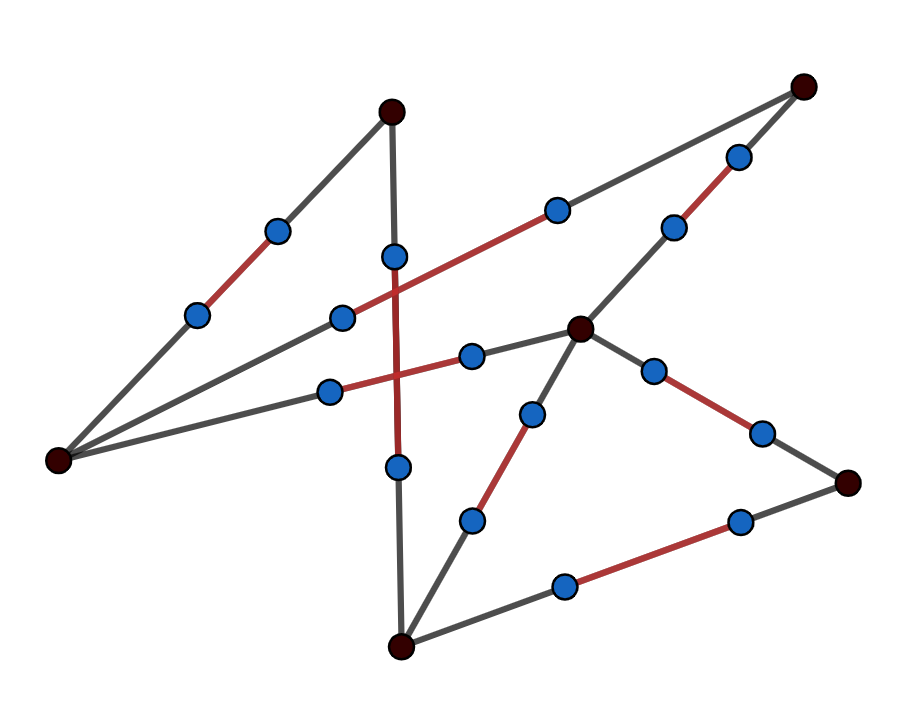}
    \caption{Graph $G$ and its $2$-subdivision}
\end{figure}
\newpage
\section{A queue number bound on the stack number}
\label{chapter_proof}
 In ~\cite{dujmovic2021stack} Dujmovi\'c $et$ $al.$ proved, that the stack number is in general not bounded by the queue number.
 \begin{theorem}
     \cite{dujmovic2021stack} For every $s \in \mathbb{N}$, there exists a graph G, such that $qn(G) \leq 4$ and $sn(G) > s$.
     \label{not_bounded}
 \end{theorem}
 In this chapter, we give our simplified version of the proof.

Let $S_a$ be the star with $a$ leaves and $H_n$ the dual of the $n \times n$ hexagonal grid as illustrated in Figure \ref{hexagonal_grid}. $G$ is the Cartesian product $S_a \square H_n$.
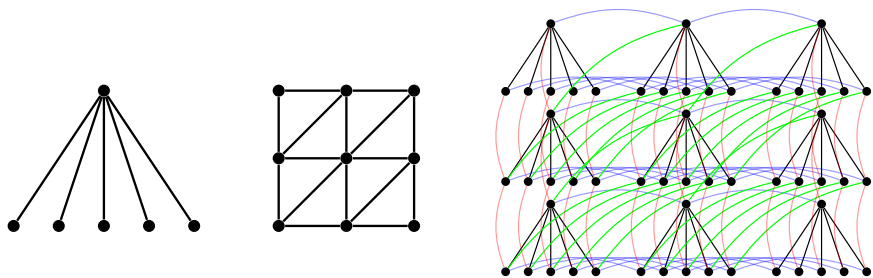
\begin{figure}[!ht]
    \centering
    \trimbox{0pt, 2pt, 0pt, 5pt}{
        \makebox[\textwidth]{
        \raisebox{0.3\height}{\scalebox{.8}{\input{pictures/star}}}
        \qquad
        \raisebox{0.3\height}{\scalebox{.8}{\input{pictures/dual_hex_grid.tex}}}
        \qquad
        \scalebox{.3}{\input{pictures/product.tex}}
        }
    }

    \caption{$S_5$, $H_3$ and their Cartesian product}

\end{figure}
\begin{lemma}
    There is a subsequence $u_1,...,u_b$ in the set $L$ of leaves  of $S_a$ of length $b \geq a^{1/2^{n^2 - 1}}$, such that for each $p \in V(H_n)$, either $(u_1, p) \prec (u_2, p) \prec ... \prec  (u_b, p)$ or $(u_1, p) \succ (u_2, p) \succ ... \succ  (u_b, p)$.
\label{erdos_lemma}
\end{lemma}
\begin{proof}
    Let $p_1, ..., p_{n^2}$ be the vertices of $H_n$. Define the starting sequence $V_1 := u_{1,1}, ..., u_{1,a_1}$ that contains all the vertices of $L$, such that $(u_{1, 1}, p_1) \prec ... \prec (u_{1,a_1}, p_1)$. By Theorem \ref{erdos-szekeres},  for $i \in \{2, ... n^2\}$, $V_{i-1}$ contains a subsequence $V_i = \{u_{i,1}, ..., u_{i, a_i}\}$, such that $a_i \geq \sqrt{a_{i-1}}$ and $(u_{i, 1}, p_i) \prec ... \prec (u_{i, a_i}, p_i)$ or $(u_{i, 1}, p_i) \succ ... \succ (u_{i, a_i}, p_i)$. An observation shows that $V_{n^2}$ is the desired sequence $u_1,...,u_b$.
\end{proof}
Define $S_b$, a star subgraph of $S_a$ with the leaves $u_1, ..., u_b$. Color every vertex $p \in H_n$ red, if $(u_1, p) \prec ... \prec (u_b, p)$. Color $p$ blue otherwise.

By Theorem \ref{hex_lemma}, we obtain a path subgraph $Q_n$ of $H_n$ on $n$ vertices, such that WLOG $(u_1, q_i) \prec ... \prec (u_b, q_i)$ for every vertex $q_i \in Q_n$.

Define a subgraph $X$ of $G$, $X := S_b \square Q_n$. Extend $\prec$ to a partial order of subsets, where for $V,W \subseteq V(X)$, $V \prec W \iff v\prec w$ for all $v \in V$ and $w \in W$.

Call $\mathcal{B}$ the set of paths defined by the leaves of $S_b$ in $X$. We say that two paths are crossing if they cross with respect to $\prec$ in any of their edges.

\begin{lemma}
    Each pair of paths $R_i, R_j \in \mathcal{B}$ are either crossing or are separated with respect to $\prec$.
\end{lemma}
\begin{proof}
    Assume $R_i$ is nested in some edge $e = (u_j, q_x)(u_j, q_y)$ of $R_j$. Since $R_i$ has an edge $(u_i, q_x)(u_i, q_y)$ nested inside $e$, we get $(u_j, q_x) \prec (u_i, q_x)$ and $(u_i, q_y) \prec (u_j, q_y)$. This is a contradiction with Lemma \ref{erdos_lemma}.
\end{proof}

    Imagine the paths in $\mathcal{B}$ as vertices. We draw a red edge between two paths if they cross, and a blue one otherwise. By Theorem \ref{ramsey_theorem}, if we have at least $R(c, d)$ paths, then there are either $c$ pairwise separated paths or $d$ pairwise crossing paths. We finish as follows.

\begin{case}
 There are $c$ pairwise separated paths in $\mathcal{B}$. 
 \label{case_1}
\end{case}
 The root of $S_b$ defines a path in $X$. Label its vertices by $s_1 \prec ... \prec s_n$. Label the paths in $\mathcal{B}$ as $R_1 \prec ... \prec R_c$. There are two subcases.
 
\begin{itemize}
     \item  $R_{\lfloor c/2 \rfloor} \prec s_{\lceil n/2 \rceil}$. There are $min\{\lfloor c/2 \rfloor, \lceil n/2 \rceil \}$ pairwise crossing edges in $X$ between vertices of $R_1, ... R_{\lfloor c/2 \rfloor}$ and $s_{\lceil n/2 \rceil}, ..., s_n$
\begin{figure}[!ht]
    \centering
    \trimbox{0pt, 5pt, 0pt, 14pt}{
    \input{pictures/separated_a}
    }   
    \caption{Case \ref{case_1}, where $R_{\lfloor c/2 \rfloor} \prec s_{\lceil n/2 \rceil}$}
\end{figure}
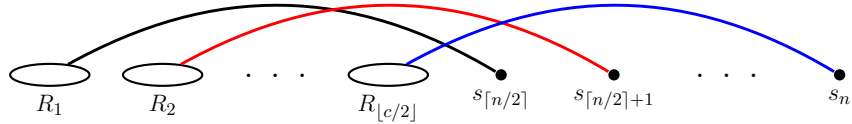 
     \item  $s_{\lceil n/2 \rceil} \prec R_{\lceil c/2 \rceil + 1}$. There are $min\{\lfloor c/2 \rfloor, \lceil n/2 \rceil \} $ pairwise crossing edges in $X$ between vertices of $R_{\lceil c/2 \rceil + 1}, ... R_c$ and $s_1, ..., s_{\lceil n/2 \rceil}$.
\begin{figure}[!ht]
    \centering
    \trimbox{0pt, 10pt, 0pt, 15pt}{
    \input{pictures/separated_b}
    }
    \caption{Case \ref{case_1}, where $s_{\lceil n/2 \rceil} \prec R_{\lceil c/2 \rceil + 1}$}
\end{figure}
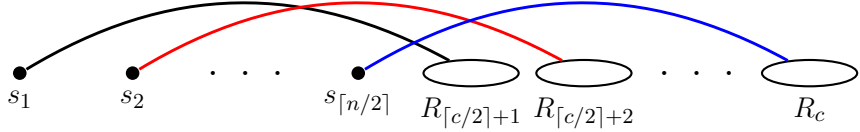 
\end{itemize}

\begin{case}
    There are $d$ pairwise crossing paths with respect to $\prec$ in $\mathcal{B}$.
\label{case_2}
\end{case}

Pick any path $P$ out of the $d$ paths. It is crossed by at least $d-1$ edges. By the pigeonhole principle $(d-1)/n$ edges cross the same edge $e$ of $P$. There are $n$ star subgraphs in $X$ defined by the vertices of $Q_n$. By the pigeonhole principle, there are at least $(d-1)/2n^2$ edges crossing $e$ whose vertices inside $e$ in $\prec$ are from the same star and their outside vertices are also from the same star. Again by the pigeonhole principle, $(d-1)/4n^2$ of those edges have their outside vertices on the same side of the edge $e$. By the choice of $Q_n$, these edges are pairwise crossing with respect to $\prec$.

\begin{figure}[!ht]
    \centering
    \input{pictures/crossing_paths}
    \caption{Case \ref{case_2}}
\end{figure}
Set $n = 2s+1$ and $a = R(2s+2, 4n^2(s+1)+1)^{2^{n^2-1}}$. We can calculate $b = R(2s+2, 4n^2(s+1)+1)$.
\begin{itemize}
    \item In Case \ref{case_1}, G has $sn(G) \geq min\{\lfloor (2s+2)/2 \rfloor, \lceil (2s+1)/2 \rceil \} > s$. 
    \item In Case \ref{case_2}, $sn(G) \geq \frac{(4n^2(s+1)+1)-1}{4n^2} > s$.
\end{itemize}
Both cases directly imply that $sn(G) > s$ by showing a subgraph of $G$ with $k$-pairwise crossing edges in a supposed $s$-stack layout of $G$, where $k > s$.

By Lemma \ref{qn_trees}, $qn(S_a) = 1$. Ordering $H_n$ row by row from the left to the right yields a $3$-strict queue layout with vertical, horizontal, and diagonal edges in three separate strict queues. By Theorem \ref{wood_product}, $qn(G) \leq 4$. This means we can construct $G$ with queue number at most $4$ and arbitrarily large stack number. Hence, we have finished the proof of Theorem \ref{not_bounded}.
\newpage
\section{Conclusion}
In Chapter \ref{chapter_stack} and Chapter \ref{chapter_queue}, we have given an overview of the stack number and the queue number. We have shown in Chapter \ref{chapter_proof}, in a simplified way, that stack number is not bounded by queue number. Simplification of our proof is in the following. In the original proof ~\cite{dujmovic2021stack}, Dujmovi\'c $et$ $al.$ use induction to show that $G$ contains a set of separated paths. To get this result, they use two more lemmas to describe the structure of the $s$-stack layout of $G$. We have rearranged this part and , instead, showed, that if $G$ does not contain $c$ separated paths then it contains $d$ pairwise crossing paths, which also leads to the conclusion. 

A lot of research has been done concerning the stack number and the queue number. However, there are still open problems and we mention a few of them here.

As mentioned in Chapter \ref{chapter_oposite}, it is unknown whether the queue number is bounded by the stack number. It is known that this problem is equivalent to deciding whether bipartite $3$-stack graphs have bounded queue number.

In Chapter \ref{chapter_proof} we have given an example of a $4$-queue graph with unbounded stack number. It is known that $1$-queue graphs accept $2$-stack layouts ~\cite{laying_queues}. It is open whether $2$ and $3$-queue graphs have bounded stack number.

Concerning graph products, it is unknown whether the strong product of $T$ and $P$ has bounded stack number for every tree $T$ and  a path $P$. It is known that this product has bounded queue number. 

\newpage
\printbibliography
\end{document}

%% file: pictures/stack_layout_example.tex
\begin{tikzpicture}[every edge/.append style={line width = 1.1pt}, every edge/.append style={line width = 1.1pt}, every node/.style={circle,thick,draw}]
    \node (A) at (0,4) {A};
    \node (B) at (2,5.5) {B};
    \node (C) at (2,4) {C};
    \node (D) at (2,2.5) {D};
    \node (E) at (4,4) {E};
    
    \node (A') at (5.5,4) {A};
    \node (B') at (7,4) {B};
    \node (C') at (8.5,4) {C};
    \node (D') at (10,4) {D};
    \node (E') at (11.5,4) {E};

    \path (A) edge (B); 
    \path (A) edge (C); 
    \path (A) edge (D); 
    \path (E) edge (B); 
    \path (E) edge (C); 
    \path (E) edge (D); 

    \path (A') edge[bend left=60] (B'); 
    \path (A') edge[bend left=60] (C'); 
    \path (A') edge[bend left=60](D'); 
    \path (E') edge[bend right=60, color=red](B'); 
    \path (E') edge[bend right=60, color=red] (C'); 
    \path (E') edge[bend right=60] (D'); 
    
\end{tikzpicture}

%% file: pictures/twist.tex
\begin{tikzpicture}[every edge/.append style={line width = 1.1pt}, every node/.style={circle,thick,draw}]
    
    \node (A')[style={circle,fill,inner sep=0pt, minimum size=0.2cm}] at (0,0) { };
    \node (B')[style={circle,fill,inner sep=0pt, minimum size=0.2cm}] at (1.5,0) { };
    \node (C')[style={circle,fill,inner sep=0pt, minimum size=0.2cm}] at (3,0) { };
    \node (D')[style={circle,fill,inner sep=0pt, minimum size=0.2cm}] at (4.5,0) { };
    \node (E')[style={circle,fill,inner sep=0pt, minimum size=0.2cm}] at (6,0) { };
    \node (F')[style={circle,fill,inner sep=0pt, minimum size=0.2cm}] at (7.5,0) { }; 

    \path (A') edge[bend left=60] (D'); 
    \path (B') edge[bend left=60, color=red] (E'); 
    \path (C') edge[bend left=60, color=blue](F'); 
    
\end{tikzpicture}

%% file: pictures/circular_layout.tex
    \begin{tikzpicture}[scale=.7, every edge/.append style={line width = 1.1pt}, every node/.style={circle,fill,inner sep=0pt, minimum size=0.2cm}]
        \node (1)[label=$w$] at (6.35, 5.87) {};
        \node (2) at (6.92, 3.33) {};
        \node (3)[label=right:$w'$] at (5.3, 1.3) {};
        \node (4) at (2.7, 1.3) {};
        \node (5) at (1.08, 3.33) {};
        \node (6) at (1.65, 5.87) {};
        \node (7) at (4.0, 7.0) {};

        \path (1) edge (2);
        \path (2) edge (7);
        \path (7) edge (3);
        \path (3) edge (6);
        \path (6) edge (4);
        \path (4) edge (5);

        \path (3) edge[color=red] (4);
        \path (4) edge[color=red] (2);
        \path (2) edge[color=red] (5);
        \path (5) edge[color=red] (1);
        \path (1) edge[color=red] (6);
        \path (6) edge[color=red] (7);

    \end{tikzpicture}

%% file: pictures/xtree.tex
    \begin{tikzpicture}[scale=.9, every edge/.append style={line width = 1.1pt}, every node/.style={circle,fill,inner sep=0pt, minimum size=0.2cm}]
        \node (8) at (0, 0) {};
        \node (9) at (1, 0) {};
        \node (10) at (1.5, 0) {};
        \node (11) at (2.5, 0) {};
        \node (12) at (3, 0) {};
        \node (13) at (4, 0) {};
        \node (14) at (4.5, 0) {};
        \node (15) at (5.5, 0) {};
        \node (4) at (0.5, 2) {};
        \node (5) at (2, 2) {};
        \node (6) at (3.5, 2) {};
        \node (7) at (5, 2) {};
        \node (2) at (1,4) {};
        \node (3) at (4.25,4) {};
        \node (1)[label=$r$] at (2.625,6) {};

        \path (1) edge (2);
        \path (1) edge (3);
        \path (2) edge (4);
        \path (2) edge (5);
        \path (3) edge (6);
        \path (3) edge (7);
        \path (2) edge (3);
        \path (4) edge (5);
        \path (5) edge (6);
        \path (6) edge (7);
        \path (4) edge (8);
        \path (4) edge (9);
        \path (5) edge (10);
        \path (5) edge (11);
        \path (6) edge (12);
        \path (6) edge (13);
        \path (7) edge (14);
        \path (7) edge (15);
        \path (8) edge (9);
        \path (9) edge (10);
        \path (10) edge (11);
        \path (11) edge (12);
        \path (12) edge (13);
        \path (13) edge (14);
        \path (14) edge (15);
    \end{tikzpicture}

%% file: pictures/queue_example.tex
\begin{tikzpicture}[every edge/.append style={line width = 1.1pt}, every node/.style={circle,thick,draw}]
    \node (A) at (0,0) {A};
    \node (B) at (0,2) {B};
    \node (C) at (2,0) {C};
    \node (D) at (2,2) {D};

    \node (A') at (6,0) {A};
    \node (B') at (7.5,0) {B};
    \node (C') at (9,0) {C};
    \node (D') at (10.5,0) {D};

    \path (A) edge (B); 
    \path (A) edge (C); 
    \path (A) edge (D); 
    \path (B) edge (C); 
    \path (B) edge (D); 
    \path (C) edge (D);
    
    \path (A') edge[bend left=60] (B'); 
    \path (A') edge[bend left=60] (C'); 
    \path (A') edge[bend left=60] (D'); 
    \path (B') edge[bend left=60, color=red] (C'); 
    \path (B') edge[bend left=60] (D'); 
    \path (C') edge[bend left=60] (D'); 
\end{tikzpicture}

%% file: pictures/rainbow.tex
\begin{tikzpicture}[scale=.8, every edge/.append style={line width = 1.1pt}, every node/.style={circle,thick,draw}]
    
    \node (A')[style={circle,fill,inner sep=0pt, minimum size=0.2cm}] at (0,0) { };
    \node (B')[style={circle,fill,inner sep=0pt, minimum size=0.2cm}] at (1.5,0) { };
    \node (C')[style={circle,fill,inner sep=0pt, minimum size=0.2cm}] at (3,0) { };
    \node (D')[style={circle,fill,inner sep=0pt, minimum size=0.2cm}] at (4.5,0) { };
    \node (E')[style={circle,fill,inner sep=0pt, minimum size=0.2cm}] at (6,0) { };
    \node (F')[style={circle,fill,inner sep=0pt, minimum size=0.2cm}] at (7.5,0) { }; 

    \path (A') edge[bend left=60] (F'); 
    \path (B') edge[bend left=60, color=red] (E'); 
    \path (C') edge[bend left=60, color=blue](D'); 
\end{tikzpicture}

%% file: pictures/leveled.tex
\begin{tikzpicture}[every edge/.append style={line width = 1.1pt}, every node/.style={circle,thick,draw}]
    \node (A)[style={circle,fill,inner sep=0pt, minimum size=0.2cm}] at (0,1) {};
    \node (B)[style={circle,fill,inner sep=0pt, minimum size=0.2cm}] at (0,3) {};
    \node (C)[style={circle,fill,inner sep=0pt, minimum size=0.2cm}] at (2,0) {};
    \node (D)[style={circle,fill,inner sep=0pt, minimum size=0.2cm}] at (2,2) {};
    \node (E)[style={circle,fill,inner sep=0pt, minimum size=0.2cm}] at (2,4) {};
    \node (F)[style={circle,fill,inner sep=0pt, minimum size=0.2cm}] at (4,2) {};
    \path (A) edge (D); 
    \path (B) edge (D); 
    \path (B) edge (E);
    \path (C) edge (F);
    \path (D) edge (F);
    \path (E) edge (F);
    \path (C) edge[bend left = 120, looseness = 2.7, color=white] (E);
\end{tikzpicture}

%% file: pictures/arched_leveled.tex
\begin{tikzpicture}[every edge/.append style={line width = 1.1pt}, every node/.style={circle,thick,draw}]
    \node (A)[style={circle,fill,inner sep=0pt, minimum size=0.2cm}] at (1,1) {};
    \node (B)[style={circle,fill,inner sep=0pt, minimum size=0.2cm}] at (1,3) {};
    \node (C)[style={circle,fill,inner sep=0pt, minimum size=0.2cm}] at (3,0) {};
    \node (D)[style={circle,fill,inner sep=0pt, minimum size=0.2cm}] at (3,2) {};
    \node (E)[style={circle,fill,inner sep=0pt, minimum size=0.2cm}] at (3,4) {};
    \node (F)[style={circle,fill,inner sep=0pt, minimum size=0.2cm}] at (5,2) {};
    
    \path (A) edge (D); 
    \path (B) edge (D); 
    \path (B) edge (E);
    \path (C) edge (F);
    \path (D) edge (F);
    \path (E) edge (F);  
    \path (A) edge[bend left = 120, color=red] (B);
    \path (C) edge[bend left = 120, looseness = 2.7, color=red] (E);
\end{tikzpicture}

%% file: pictures/nested_leveled.tex
\begin{tikzpicture}[scale=1.15, every edge/.append style={line width = 1.1pt}, every node/.style={circle,fill,inner sep=0pt, minimum size=0.2cm}]
        \node (1)[label=below:$s_1$] at (0,0) {};
        \node (2)[label=below:$s_2$] at (1,0) {};
        \node (3)[label=below:$r_2$] at (2,0) {};
        \node (4)[label=below:$r_1$] at (3,0) {};

        \path (1)edge[bend left=40] (4);
        \path (2)edge[color=red,bend left=40] (3);
    \end{tikzpicture}

%% file: pictures/crossed_leveled.tex
\begin{tikzpicture}[scale=1.15, every edge/.append style={line width = 1.1pt}, every node/.style={circle,fill,inner sep=0pt, minimum size=0.2cm}]
        \node (1)[label=left:$s_1$] at (0,.5) {};
        \node (2)[label=left:$s_2$] at (0,1.5) {};
        \node (3)[label=right:$r_2$] at (2,.5) {};
        \node (4)[label=right:$r_1$] at (2,1.5) {};

        \draw (0, 0)[label=below:$l_i$] -- (0, 2);
        \draw (2, 0)[label=below:$l_{i+1}$] -- (2, 2);

        \path (1)edge[] (4);
        \path (2)edge[color=red] (3);
    \end{tikzpicture}

%% file: pictures/k44layout.tex
    \begin{tikzpicture}[every edge/.append style={line width = 1.1pt}, every node/.style={circle,fill,inner sep=0pt, minimum size=0.2cm}]
        \node (1)[label=below:$a_1$] at (0,0) {};
        \node (2)[label=below:$a_2$] at (1,0) {};
        \node (3)[label=below:$b_1$, color=gray] at (2,0) {};
        \node (4)[label=below:$b_2$, color=gray] at (3,0) {};
        \node (5)[label=below:$b_3$, color=gray] at (4,0) {};
        \node (6)[label=below:$b_4$, color=gray] at (5,0) {};
        \node (7)[label=below:$a_3$] at (6,0) {};
        \node (8)[label=below:$a_4$] at (7,0) {};

        \path (1) edge[bend left=40] (3);
        \path (1) edge[bend left=40] (4);
        \path (1) edge[bend left=40] (5);
        \path (1) edge[bend left=40] (6);
        
        \path (8) edge[bend right=40] (3);
        \path (8) edge[bend right=40] (4);
        \path (8) edge[bend right=40] (5);
        \path (8) edge[bend right=40] (6);
        
    \end{tikzpicture}

%% file: pictures/c4a.tex
\begin{tikzpicture}[scale=.8, every edge/.append style={line width = 1.1pt}, every node/.style={circle,fill,inner sep=0pt, minimum size=0.2cm}]
        \node (A) at (1, 0) {};
        \node (B) at (2.5, 0) {};
        \node (C) at (1, 1.5) {};
        \node (D) at (2.5, 1.5) {};
        
        \node (H1) at (0, 0) {};
        \node (H2) at (3.5, 0) {};
        \node (H3) at (1.75, 3) {};

        \path (A) edge[color=red] (B);
        \path (B) edge[color=red] (D);
        \path (D) edge[color=red] (C);
        \path (C) edge[color=red] (A);

        \path (H1) edge (A);
        \path (H1) edge (C);
        \path (H2) edge (B);
        \path (H2) edge (D);
        \path (H3) edge (C);
        \path (H3) edge (D);
        
    \end{tikzpicture}

%% file: pictures/c4b.tex
    \begin{tikzpicture}[scale = .75, every edge/.append style={line width = 1.1pt}, every node/.style={circle,thick,draw}]
 
    \node (A) at (0,0) {$u_1$};
    \node (B) at (2.5,0) {$u_2$};
    \node (D) at (5,0) {$u_4$};
    \node (C) at (7.5,0) {$u_3$};
    \node (P)[label=below:$p$, style={circle,fill,inner sep=0pt, minimum size=0.2cm}] at (1.25,0) {};

    \path (A) edge[bend left=60] (B); 
    \path (B) edge[bend left=60] (C); 
    \path (D) edge[bend left=60] (C); 
    \path (A) edge[bend left=60] (D);
    \path (P) edge[bend right=60, color=red] (A);
    \path (P) edge[bend left=60, color=red] (D);
\end{tikzpicture}

%% file: pictures/arched_cycle_a.tex
\begin{tikzpicture}[every edge/.append style={line width = 1.1pt}, every node/.style={circle,fill,inner sep=0pt, minimum size=0.2cm}]
    
    \node (1)[label=below:$u_1$] at (0,0) {};
    \node (6)[label=above:$u_7$] at (0,1) {};
    \node (2)[label=below:$u_2$] at (1.5,0) {};
    \node (5)[label=above:$u_6$] at (1.5,1) {};
    \node (3)[label=below:$u_3$] at (3,0) {};
    \node (4)[label=above:$u_5$] at (3,1) {};
    \node (7)[label=below:$u_4$] at (4.5,0.5) {};

    \path (1) edge (6);
    \path (6) edge (5);
    \path (5) edge (4);
    \path (4) edge (7);
    \path (7) edge (3);
    \path (3) edge (2);
    \path (2) edge (1);
\end{tikzpicture}

%% file: pictures/arched_cycle_b.tex
\begin{tikzpicture}[every edge/.append style={line width = 1.1pt}, every node/.style={circle,fill,inner sep=0pt, minimum size=0.2cm}]
    \node (A)[label=below:$u_1$] at (0,0.5) {};
    \node (B)[label=above:$u_2$] at (1.5,1) {};
    \node (C)[label=below:$u_6$] at (1.5,0) {};
    \node (D)[label=above:$u_3$] at (3,1) {};
    \node (E)[label=below:$u_5$] at (3,0) {};
    \node (F)[label=below:$u_4$] at (4.5,0.5) {};

    \path (A) edge (B);
    \path (A) edge (C);
    \path (C) edge (E);
    \path (B) edge (D);
    \path (D) edge (F);
    \path (E) edge (F);
\end{tikzpicture}

%% file: pictures/cartesian.tex
\begin{tikzpicture}[every edge/.append style={line width = 1.1pt}, every node/.style={circle,fill,inner sep=0pt, minimum size=0.2cm}]
    
    \node (1) at (0,0) {};
    \node (2) at (1,0) {};
    \node (3) at (2,0) {};
    \node (4) at (3,0) {};

    \node (5) at (5,0) {};
    \node (6) at (6,0) {};
    \node (7) at (7,0) {};
    \node (8) at (8,0) {};

    \node (9) at (10,0) {};
    \node (10) at (11,0) {};
    \node (11) at (12,0) {};
    \node (12) at (13,0) {};

    \path (1) edge[bend left=50, opacity=.8](2);
    \path (1) edge[bend left=50, opacity=.8](3);
    \path (1) edge[bend left=50, opacity=.8](4);

    \path (5) edge[bend left=50, opacity=.8](6);
    \path (5) edge[bend left=50, opacity=.8](7);
    \path (5) edge[bend left=50, opacity=.8](8);

    \path (9) edge[bend left=50, opacity=.8](10);
    \path (9) edge[bend left=50, opacity=.8](11);
    \path (9) edge[bend left=50, opacity=.8](12);

    \path (1) edge[bend left=50, opacity=.7, color=red](5);
    \path (5) edge[bend left=50, opacity=.7, color=red](9);

    \path (2) edge[bend left=50, opacity=.3, color=red](6);
    \path (6) edge[bend left=50, opacity=.3, color=red](10);
    
    \path (3) edge[bend left=50, opacity=.3, color=red](7);
    \path (7) edge[bend left=50, opacity=.3, color=red](11);
    
    \path (4) edge[bend left=50, opacity=.7, color=red](8);
    \path (8) edge[bend left=50, opacity=.7, color=red](12);

\end{tikzpicture}

%% file: pictures/star.tex
\begin{tikzpicture}[scale=0.75, every edge/.append style={line width = 1.1pt}, every node/.style={circle,fill,inner sep=0pt, minimum size=0.2cm}]
    \node (S) at (2,3) { };
    \node (A) at (0,0) { };
    \node (B) at (1,0) { };
    \node (C) at (2,0) { };
    \node (D) at (3,0) { };
    \node (E) at (4,0) { };    
    \path (S) edge (A); 
    \path (S) edge (B); 
    \path (S) edge (C);
    \path (S) edge (D);
    \path (S) edge (E);
    
\end{tikzpicture}

%% file: pictures/dual_hex_grid.tex
\begin{tikzpicture}[scale=0.75, every edge/.append style={line width = 1.1pt}, every node/.style={circle,fill,inner sep=0pt, minimum size=0.2cm}]   
    \node (1) at (0,0) { };
    \node (2) at (1.5,0) { };
    \node (3) at (3,0) { };
    \node (4) at (0,1.5) { };
    \node (5) at (1.5,1.5) { };
    \node (6) at (3,1.5) { }; 
    \node (7) at (0,3) { };
    \node (8) at (1.5,3) { };
    \node (9) at (3,3) { };
    
    \path (1) edge (2); 
    \path (2) edge (3); 
    \path (4) edge (5); 
    \path (5) edge (6); 
    \path (7) edge (8); 
    \path (8) edge (9); 
    \path (1) edge (5); 
    \path (2) edge (6); 
    \path (4) edge (8); 
    \path (5) edge (9); 
    \path (1) edge (4); 
    \path (4) edge (7);
    \path (2) edge (5); 
    \path (5) edge (8); 
    \path (3) edge (6); 
    \path (6) edge (9); 
\end{tikzpicture}

%% file: pictures/product.tex
    
\begin{tikzpicture}[every edge/.append style={line width = 1.5pt}, every node/.style={circle,fill=black}]
    
    \node (S1) at (2,3) { };
    \node (A1) at (0,0) { };
    \node (B1) at (1,0) { };
    \node (C1) at (2,0) { };
    \node (D1) at (3,0) { };
    \node (E1) at (4,0) { }; 
    \path (S1) edge (A1); 
    \path (S1) edge (B1); 
    \path (S1) edge (C1);
    \path (S1) edge (D1);
    \path (S1) edge (E1);
    
    \node (S2) at (8,3) { };
    \node (A2) at (6,0) { };
    \node (B2) at (7,0) { };
    \node (C2) at (8,0) { };
    \node (D2) at (9,0) { };
    \node (E2) at (10,0) { }; 
    \path (S2) edge (A2); 
    \path (S2) edge (B2); 
    \path (S2) edge (C2);
    \path (S2) edge (D2);
    \path (S2) edge (E2);

    \node (S3) at (14,3) { };
    \node (A3) at (12,0) { };
    \node (B3) at (13,0) { };
    \node (C3) at (14,0) { };
    \node (D3) at (15,0) { };
    \node (E3) at (16,0) { }; 
    \path (S3) edge (A3); 
    \path (S3) edge (B3); 
    \path (S3) edge (C3);
    \path (S3) edge (D3);
    \path (S3) edge (E3);

    \node (S4) at (2,7) { };
    \node (A4) at (0,4) { };
    \node (B4) at (1,4) { };
    \node (C4) at (2,4) { };
    \node (D4) at (3,4) { };
    \node (E4) at (4,4) { }; 
    \path (S4) edge (A4); 
    \path (S4) edge (B4); 
    \path (S4) edge (C4);
    \path (S4) edge (D4);
    \path (S4) edge (E4);
    
    \node (S5) at (8,7) { };
    \node (A5) at (6,4) { };
    \node (B5) at (7,4) { };
    \node (C5) at (8,4) { };
    \node (D5) at (9,4) { };
    \node (E5) at (10,4) { }; 
    \path (S5) edge (A5); 
    \path (S5) edge (B5); 
    \path (S5) edge (C5);
    \path (S5) edge (D5);
    \path (S5) edge (E5);

    \node (S6) at (14,7) { };
    \node (A6) at (12,4) { };
    \node (B6) at (13,4) { };
    \node (C6) at (14,4) { };
    \node (D6) at (15,4) { };
    \node (E6) at (16,4) { }; 
    \path (S6) edge (A6); 
    \path (S6) edge (B6); 
    \path (S6) edge (C6);
    \path (S6) edge (D6);
    \path (S6) edge (E6);

    \node (S7) at (2,11) { };
    \node (A7) at (0,8) { };
    \node (B7) at (1,8) { };
    \node (C7) at (2,8) { };
    \node (D7) at (3,8) { };
    \node (E7) at (4,8) { }; 
    \path (S7) edge (A7); 
    \path (S7) edge (B7); 
    \path (S7) edge (C7);
    \path (S7) edge (D7);
    \path (S7) edge (E7);
    
    \node (S8) at (8,11) { };
    \node (A8) at (6,8) { };
    \node (B8) at (7,8) { };
    \node (C8) at (8,8) { };
    \node (D8) at (9,8) { };
    \node (E8) at (10,8) { }; 
    \path (S8) edge (A8); 
    \path (S8) edge (B8); 
    \path (S8) edge (C8);
    \path (S8) edge (D8);
    \path (S8) edge (E8);

    \node (S9) at (14,11) { };
    \node (A9) at (12,8) { };
    \node (B9) at (13,8) { };
    \node (C9) at (14,8) { };
    \node (D9) at (15,8) { };
    \node (E9) at (16,8) { }; 
    \path (S9) edge (A9); 
    \path (S9) edge (B9); 
    \path (S9) edge (C9);
    \path (S9) edge (D9);
    \path (S9) edge (E9);

    \path (S1) edge[color=blue, opacity=0.4, bend left=20] (S2); 
    \path (S2) edge[color=blue, opacity=0.4, bend left=20] (S3); 
    \path (S4) edge[color=blue, opacity=0.4, bend left=20] (S5); 
    \path (S5) edge[color=blue, opacity=0.4, bend left=20] (S6); 
    \path (S7) edge[color=blue, opacity=0.4, bend left=20] (S8); 
    \path (S8) edge[color=blue, opacity=0.4, bend left=20] (S9); 
    \path (A1) edge[color=blue, opacity=0.4, bend left=20] (A2); 
    \path (A2) edge[color=blue, opacity=0.4, bend left=20] (A3); 
    \path (A4) edge[color=blue, opacity=0.4, bend left=20] (A5); 
    \path (A5) edge[color=blue, opacity=0.4, bend left=20] (A6); 
    \path (A7) edge[color=blue, opacity=0.4, bend left=20] (A8); 
    \path (A8) edge[color=blue, opacity=0.4, bend left=20] (A9); 
    \path (B1) edge[color=blue, opacity=0.4, bend left=20] (B2); 
    \path (B2) edge[color=blue, opacity=0.4, bend left=20] (B3); 
    \path (B4) edge[color=blue, opacity=0.4, bend left=20] (B5); 
    \path (B5) edge[color=blue, opacity=0.4, bend left=20] (B6); 
    \path (B7) edge[color=blue, opacity=0.4, bend left=20] (B8); 
    \path (B8) edge[color=blue, opacity=0.4, bend left=20] (B9); 
    \path (C1) edge[color=blue, opacity=0.4, bend left=20] (C2); 
    \path (C2) edge[color=blue, opacity=0.4, bend left=20] (C3); 
    \path (C4) edge[color=blue, opacity=0.4, bend left=20] (C5); 
    \path (C5) edge[color=blue, opacity=0.4, bend left=20] (C6); 
    \path (C7) edge[color=blue, opacity=0.4, bend left=20] (C8); 
    \path (C8) edge[color=blue, opacity=0.4, bend left=20] (C9); 
    \path (D1) edge[color=blue, opacity=0.4, bend left=20] (D2); 
    \path (D2) edge[color=blue, opacity=0.4, bend left=20] (D3); 
    \path (D4) edge[color=blue, opacity=0.4, bend left=20] (D5); 
    \path (D5) edge[color=blue, opacity=0.4, bend left=20] (D6); 
    \path (D7) edge[color=blue, opacity=0.4, bend left=20] (D8); 
    \path (D8) edge[color=blue, opacity=0.4, bend left=20] (D9); 
    \path (E1) edge[color=blue, opacity=0.4, bend left=20] (E2); 
    \path (E2) edge[color=blue, opacity=0.4, bend left=20] (E3); 
    \path (E4) edge[color=blue, opacity=0.4, bend left=20] (E5); 
    \path (E5) edge[color=blue, opacity=0.4, bend left=20] (E6); 
    \path (E7) edge[color=blue, opacity=0.4, bend left=20] (E8); 
    \path (E8) edge[color=blue, opacity=0.4, bend left=20] (E9);
    \path (A1) edge[color=green,bend left=20] (A5); 
    \path (A2) edge[color=green,bend left=20] (A6); 
    \path (A4) edge[color=green,bend left=20] (A8); 
    \path (A5) edge[color=green,bend left=20] (A9); 
    \path (B1) edge[color=green,bend left=20] (B5); 
    \path (B2) edge[color=green,bend left=20] (B6); 
    \path (B4) edge[color=green,bend left=20] (B8); 
    \path (B5) edge[color=green,bend left=20] (B9); 
    \path (C1) edge[color=green,bend left=20] (C5); 
    \path (C2) edge[color=green,bend left=20] (C6); 
    \path (C4) edge[color=green,bend left=20] (C8); 
    \path (C5) edge[color=green,bend left=20] (C9); 
    \path (D1) edge[color=green,bend left=20] (D5); 
    \path (D2) edge[color=green,bend left=20] (D6); 
    \path (D4) edge[color=green,bend left=20] (D8); 
    \path (D5) edge[color=green,bend left=20] (D9); 
    \path (E1) edge[color=green,bend left=20] (E5); 
    \path (E2) edge[color=green,bend left=20] (E6); 
    \path (E4) edge[color=green,bend left=20] (E8); 
    \path (E5) edge[color=green,bend left=20] (E9); 
    \path (S1) edge[color=green,bend left=20] (S5); 
    \path (S2) edge[color=green,bend left=20] (S6); 
    \path (S4) edge[color=green,bend left=20] (S8); 
    \path (S5) edge[color=green,bend left=20] (S9);
    \path (A1) edge[color=red, opacity=0.4, bend left=20] (A4); 
    \path (A4) edge[color=red, opacity=0.4, bend left=20] (A7);
    \path (A2) edge[color=red, opacity=0.4, bend left=20] (A5); 
    \path (A5) edge[color=red, opacity=0.4, bend left=20] (A8); 
    \path (A3) edge[color=red, opacity=0.4, bend left=20] (A6); 
    \path (A6) edge[color=red, opacity=0.4, bend left=20] (A9); 
    \path (B1) edge[color=red, opacity=0.4, bend left=20] (B4); 
    \path (B4) edge[color=red, opacity=0.4, bend left=20] (B7);
    \path (B2) edge[color=red, opacity=0.4, bend left=20] (B5); 
    \path (B5) edge[color=red, opacity=0.4, bend left=20] (B8); 
    \path (B3) edge[color=red, opacity=0.4, bend left=20] (B6); 
    \path (B6) edge[color=red, opacity=0.4, bend left=20] (B9); 
    \path (C1) edge[color=red, opacity=0.4, bend left=20] (C4); 
    \path (C4) edge[color=red, opacity=0.4, bend left=20] (C7);
    \path (C2) edge[color=red, opacity=0.4, bend left=20] (C5); 
    \path (C5) edge[color=red, opacity=0.4, bend left=20] (C8); 
    \path (C3) edge[color=red, opacity=0.4, bend left=20] (C6); 
    \path (C6) edge[color=red, opacity=0.4, bend left=20] (C9); 
    \path (D1) edge[color=red, opacity=0.4, bend left=20] (D4); 
    \path (D4) edge[color=red, opacity=0.4, bend left=20] (D7);
    \path (D2) edge[color=red, opacity=0.4, bend left=20] (D5); 
    \path (D5) edge[color=red, opacity=0.4, bend left=20] (D8); 
    \path (D3) edge[color=red, opacity=0.4, bend left=20] (D6); 
    \path (D6) edge[color=red, opacity=0.4, bend left=20] (D9); 
    \path (E1) edge[color=red, opacity=0.4, bend left=20] (E4); 
    \path (E4) edge[color=red, opacity=0.4, bend left=20] (E7);
    \path (E2) edge[color=red, opacity=0.4, bend left=20] (E5); 
    \path (E5) edge[color=red, opacity=0.4, bend left=20] (E8); 
    \path (E3) edge[color=red, opacity=0.4, bend left=20] (E6); 
    \path (E6) edge[color=red, opacity=0.4, bend left=20] (E9); 
    \path (S1) edge[color=red, opacity=0.4, bend left=20] (S4); 
    \path (S4) edge[color=red, opacity=0.4, bend left=20] (S7);
    \path (S2) edge[color=red, opacity=0.4, bend left=20] (S5); 
    \path (S5) edge[color=red, opacity=0.4, bend left=20] (S8); 
    \path (S3) edge[color=red, opacity=0.4, bend left=20] (S6); 
    \path (S6) edge[color=red, opacity=0.4, bend left=20] (S9); 

\end{tikzpicture}

%% file: pictures/separated_a.tex
\begin{tikzpicture}[every edge/.append style={line width = 1.1pt}, every node/.style={thick, scale=0.75}]
    
    \node [style=ellipse, minimum width=40pt, label=below:$R_1$, draw](A) at (0,0) {};
    \node [style=ellipse, minimum width=40pt, label=below:$R_2$, draw](B) at (1.5,0) {};
    \node (C) [ellipse, minimum width=100pt, color=white, text = black]at (3,0) {\LARGE . . .};
    \node [ellipse, minimum width=40pt, label=below:$R_{\lfloor c/2 \rfloor}$, draw](D) at (4.5,0) {};
    \node [style={circle,fill,inner sep=0pt, minimum size=0.2cm}, label=below:$s_{\lceil n/2 \rceil}$](A') at (6,0) { };
    \node [style={circle,fill,inner sep=0pt, minimum size=0.2cm}, label=below:$s_{\lceil n/2 \rceil + 1}$](B') at (7.5,0) { };
    \node [ style={circle,fill,inner sep=0pt, minimum size=0.2cm}, color = white, text = black](C') at (9,0) {\LARGE . . .};
    \node [style={circle,fill,inner sep=0pt, minimum size=0.2cm}, label=below:$s_n$](D') at (10.5,0) { };
    
    \path (A) edge[bend left=30] (A'); 
    \path (B) edge[bend left=30, color=red] (B'); 
    \path (D) edge[bend left=30, color=blue](D'); 
    
\end{tikzpicture}

%% file: pictures/separated_b.tex
\begin{tikzpicture}[every edge/.append style={line width = 1.1pt}, every node/.style={thick, scale=.9}]
    
    \node [style={circle,fill,inner sep=0pt, minimum size=0.2cm}, label=below:$s_1$](A) at (0,0) {};
    \node [style={circle,fill,inner sep=0pt, minimum size=0.2cm}, label=below:$s_2$](B) at (1.5,0) {};
    \node [style={circle,fill,inner sep=0pt, minimum size=0.2cm}, color = white, text = black](C) at (3,0) {\LARGE . . .};
    \node [style={circle,fill,inner sep=0pt, minimum size=0.2cm}, label=below:$s_{\lceil n/2 \rceil}$](D) at (4.5,0) {};
    \node [style=draw, ellipse, minimum width=40pt, label=below:$R_{\lceil c/2 \rceil + 1}$](A') at (6,0) { };
    \node [style=draw, ellipse, minimum width=40pt, label=below:$R_{\lceil c/2 \rceil + 2}$](B') at (7.5,0) { };
    \node [ellipse, minimum width=100pt, color=white, text = black](C') at (9,0) {\LARGE . . .};
    \node [draw, ellipse, minimum width=40pt, label=below:$R_c$](D') at (10.5,0) { };
    
    \path (A) edge[bend left=30] (A'); 
    \path (B) edge[bend left=30, color=red] (B'); 
    \path (D) edge[bend left=30, color=blue](D'); 
    
\end{tikzpicture}

%% file: pictures/crossing_paths.tex
   \begin{tikzpicture}[scale=.8, every edge/.append style={line width = 1.1pt}]
        \node (w) [style={circle,fill,inner sep=0pt, minimum size=0.2cm}] at (0, 0) { };
        \node (ww)[style={circle,fill,inner sep=0pt, minimum size=0.2cm}] at (6, 0) { };
        \node (x)[style={opacity=0, label}, label=below:$e$] at (1, 1.5) {};
        \path (w) edge[bend left = 40, color=red] (ww);
        
        \node (S1)[style={circle,fill,inner sep=0pt, minimum size=0.2cm}] at (2,2) { };
        \node (A1)[style={circle,fill,inner sep=0pt, minimum size=0.2cm}] at (2,0) { };
        \node (B1)[style={circle,fill,inner sep=0pt, minimum size=0.2cm}] at (2.5,0) { };
        \node (C1)[style={circle,fill,inner sep=0pt, minimum size=0.2cm}] at (3,0) { };
        \node (D1)[style={circle,fill,inner sep=0pt, minimum size=0.2cm}] at (3.5,0) { };
        \node (E1)[style={circle,fill,inner sep=0pt, minimum size=0.2cm}] at (4,0) { };    
        \path (S1) edge[opacity=0.5] (A1); 
        \path (S1) edge[opacity=0.5] (B1); 
        \path (S1) edge[opacity=0.5] (C1);
        \path (S1) edge[opacity=0.5] (D1);
        \path (S1) edge[opacity=0.5] (E1);

        \node (S2)[style={circle,fill,inner sep=0pt, minimum size=0.2cm}] at (10,2) { };
        \node (A2)[style={circle,fill,inner sep=0pt, minimum size=0.2cm}] at (8,0) { };
        \node (B2)[style={circle,fill,inner sep=0pt, minimum size=0.2cm}] at (8.5,0) { };
        \node (C2)[style={circle,fill,inner sep=0pt, minimum size=0.2cm}] at (9,0) { };
        \node (D2)[style={circle,fill,inner sep=0pt, minimum size=0.2cm}] at (9.5,0) { };
        \node (E2)[style={circle,fill,inner sep=0pt, minimum size=0.2cm}] at (10,0) { };    
        \path (S2) edge[opacity=0.5] (A2); 
        \path (S2) edge[opacity=0.5] (B2); 
        \path (S2) edge[opacity=0.5] (C2);
        \path (S2) edge[opacity=0.5] (D2);
        \path (S2) edge[opacity=0.5] (E2);

        \path (A1) edge[color=blue, bend left = 40, opacity=60] (A2);
        \path (B1) edge[color=blue, bend left = 40, opacity=60] (B2);
        \path (C1) edge[color=blue, bend left = 40, opacity=60] (C2);
        \path (D1) edge[color=blue, bend left = 40, opacity=60] (D2);
        \path (E1) edge[color=blue, bend left = 40, opacity=60] (E2);
    \end{tikzpicture}